\documentclass[prl,twocolumn,superscriptaddress,showpacs]{revtex4-2}
\usepackage[dvipsnames]{xcolor}
\usepackage{amsmath}
\usepackage{units,xspace}
\usepackage{graphicx}
\usepackage{array}
\usepackage{url,hyperref}
\usepackage{multirow}
\usepackage[utf8]{inputenc}

\newcommand{\fb}{\mbox{fb$^{-1}$}\xspace}

\newcommand{\gd}{\ensuremath{g_{\textrm D}}\xspace}

\newcommand{\geant}{\textsc{Geant4}\xspace}

\DeclareGraphicsExtensions{.pdf,.png,.jpg,.jpeg}
\graphicspath{{./}{./fig/}}

\begin{document}


\sloppy



\newpage

\title{Search for Highly-Ionizing Particles in  \\ $pp$ Collisions at the LHC's Run-1 Using the Prototype MoEDAL Detector }

\author{B.~Acharya}
\altaffiliation[Also at ]{Int. Centre for Theoretical Physics, Trieste, Italy}   
\affiliation{Theoretical Particle Physics \& Cosmology Group, Physics Dept., King's College London, UK}

\author{J.~Alexandre}
\affiliation{Theoretical Particle Physics \& Cosmology Group, Physics Dept., King's College London, UK}


\author{P.~Benes}
\affiliation{IEAP, Czech Technical University in Prague, Czech~Republic}

\author{B.~Bergmann}
\affiliation{IEAP, Czech Technical University in Prague, Czech~Republic}


\author{S. Bertolucci}
\affiliation{INFN, Section of Bologna, Bologna, Italy}

\author{A.~Bevan}
\affiliation{School of Physics and Astronomy, Queen Mary University of London, UK}

\author{R.~Bhattacharya}
\altaffiliation{CAPSS,
Department of Physics
Bose Institute, Kolkata, India}
\affiliation{INFN, Section of Bologna, Bologna, Italy}

\author{H.~Branzas}
\affiliation{Institute of Space Science, Bucharest - M\u{a}gurele, Romania}

\author{P. Burian}
\affiliation{IEAP, Czech Technical University in Prague, Czech~Republic}

\author{M.~Campbell}
\affiliation{Experimental Physics Department, CERN, Geneva, Switzerland}

\author{S.~Cecchini}
\affiliation{INFN, Section of Bologna, Bologna, Italy}

\author{Y.~M.~Cho}
\affiliation{Center for Quantum Spacetime, Sogang University, Seoul, Korea} 

\author{M.~de~Montigny}
\affiliation{Physics Department, University of Alberta, Edmonton, Alberta, Canada}

\author{A.~De~Roeck}
\affiliation{Experimental Physics Department, CERN, Geneva, Switzerland}

\author{J.~R.~Ellis}
\altaffiliation[Also at ]{National Institute of Chemical Physics \& Biophysics, Tallinn, Estonia} 
\affiliation{Theoretical Particle Physics \& Cosmology Group, Physics Dept., King's College London, UK}
\affiliation{Theoretical Physics Department, CERN, Geneva, Switzerland}

\author{M.~El~Sawy}
\altaffiliation[Also at ]{Dept. of Physics, Faculty of Science, Beni-Suef University, Egypt} 
\affiliation{Experimental Physics Department, CERN, Geneva, Switzerland}

\author{M.~Fairbairn}
\affiliation{Theoretical Particle Physics \& Cosmology Group, Physics Dept., King's College London, UK}

\author{D.~Felea}
\affiliation{Institute of Space Science, Bucharest - M\u{a}gurele, Romania}

\author{M.~Frank}
\affiliation{Department of Physics, Concordia University, Montr\'{e}al, Qu\'{e}bec,  Canada}

\author{J.~Hays}
\affiliation{School of Physics and Astronomy, Queen Mary University of London, UK}

\author{A.~M.~Hirt}
\affiliation{Department of Earth Sciences, Swiss Federal Institute of Technology, Zurich, Switzerland}

\author{P. Q.~Hung}
\affiliation{Department of Physics, University of Virginia, Charlottesville, Virginia, USA}

\author{J.~Janecek}
\affiliation{IEAP, Czech Technical University in Prague, Czech~Republic}

\author{M.~Kalliokoski}
\affiliation{Helsinki Institute of Physics, University of Helsinki, Helsinki, Finland}


\author{A.~Korzenev}
\affiliation{D\'epartement de Physique Nucl\'eaire et Corpusculaire, Universit\'e de Gen\`eve, Geneva, Switzerland}

\author{D.~H.~Lacarr\`ere}
\affiliation{Experimental Physics Department, CERN, Geneva, Switzerland}


\author{C.~Leroy}
\affiliation{D\'{e}partement de Physique, Universit\'{e} de Montr\'{e}al, Qu\'{e}bec, Canada}

\author{G.~Levi} 
\affiliation{INFN, Section of Bologna \& Department of Physics \& Astronomy, University of Bologna, Italy}

\author{A.~Lionti}
\affiliation{D\'epartement de Physique Nucl\'eaire et Corpusculaire, Universit\'e de Gen\`eve, Geneva, Switzerland}


\author{A.~Margiotta}
\affiliation{INFN, Section of Bologna \& Department of Physics \& Astronomy, University of Bologna, Italy}

\author{R.~Maselek}
\affiliation{Institute of Theoretical Physics, University of Warsaw, Warsaw, Poland}

\author{A. Maulik}
\affiliation{INFN, Section of Bologna, Bologna, Italy}
\affiliation{Physics Department, University of Alberta, Edmonton, Alberta, Canada}

\author{N.~Mauri}
\affiliation{INFN, Section of Bologna \& Department of Physics \& Astronomy, University of Bologna, Italy}

\author{N.~E.~Mavromatos}
\altaffiliation[Also at ]{Department of Physics, School of Applied Mathematical and Physical Sciences, 
National Technical University of Athens, Athens, Greece} 
\affiliation{Theoretical Particle Physics \& Cosmology Group, Physics Dept., King's College London, UK}

\author{E.~Musumeci}
\affiliation{IFIC, Universitat de Val\`{e}ncia - CSIC, Valencia, Spain}

\author{M.~Mieskolainen}
\affiliation{Physics Department, University of Helsinki, Helsinki, Finland}

\author{L.~Millward}
\affiliation{School of Physics and Astronomy, Queen Mary University of London, UK}

\author{V.~A.~Mitsou}
\affiliation{IFIC, Universitat de Val\`{e}ncia - CSIC, Valencia, Spain}

\author{R.~Orava}
\affiliation{Physics Department, University of Helsinki, Helsinki, Finland}

\author{I.~Ostrovskiy}
\affiliation{Department of Physics and Astronomy, University of Alabama, Tuscaloosa, Alabama, USA}

\author{P.-P. Ouimet}
\affiliation{Physics Department, University of Regina, Regina, Saskatchewan, Canada}
  
\author{J.~Papavassiliou}
\affiliation{IFIC, Universitat de Val\`{e}ncia - CSIC, Valencia, Spain}

\author{B.~Parker}
\affiliation{Institute for Research in Schools, Canterbury, UK}

\author{L.~Patrizii}
\email[Corresponding author: ]{Laura.Patrizii@bo.infn.it}
\affiliation{INFN, Section of Bologna, Bologna, Italy}

\author{G.~E.~P\u{a}v\u{a}la\c{s}}
\affiliation{Institute of Space Science, Bucharest - M\u{a}gurele, Romania}

\author{J.~L.~Pinfold}
\email[Corresponding author: ]{jpinfold@ualberta.ca}
\affiliation{Physics Department, University of Alberta, Edmonton, Alberta, Canada}

\author{L.~A.~Popa}
\affiliation{Institute of Space Science, Bucharest - M\u{a}gurele, Romania}

\author{V.~Popa}
\affiliation{Institute of Space Science, Bucharest - M\u{a}gurele, Romania}

\author{M.~Pozzato}
\affiliation{INFN, Section of Bologna, Bologna, Italy}

\author{S.~Pospisil}
\affiliation{IEAP, Czech Technical University in Prague, Czech~Republic}

\author{A.~Rajantie}
\affiliation{Department of Physics, Imperial College London, UK}

\author{R.~Ruiz~de~Austri}
\affiliation{IFIC, Universitat de Val\`{e}ncia - CSIC, Valencia, Spain}

\author{Z.~Sahnoun}
\affiliation{INFN, Section of Bologna, Bologna, Italy}
\affiliation{Research Centre for Astronomy, Astrophysics and Geophysics, Algiers, Algeria}

\author{M.~Sakellariadou}
\affiliation{Theoretical Particle Physics \& Cosmology Group, Physics Dept., King's College London, UK}

\author{K.~Sakurai}
\affiliation{Institute of Theoretical Physics, University of Warsaw, Warsaw, Poland}

\author{A.~Santra}
\affiliation{IFIC, Universitat de Val\`{e}ncia - CSIC, Valencia, Spain}

\author{S.~Sarkar}
\affiliation{Theoretical Particle Physics \& Cosmology Group, Physics Dept., King's College London, UK}

\author{G.~Semenoff}
\affiliation{Department of Physics, University of British Columbia, Vancouver, British Columbia, Canada}

\author{A.~Shaa}
\affiliation{Physics Department, University of Alberta, Edmonton, Alberta, Canada}

\author{G.~Sirri}
\affiliation{INFN, Section of Bologna, Bologna, Italy}

\author{K.~Sliwa}
\affiliation{Department of Physics and Astronomy, Tufts University, Medford, Massachusetts, USA}

\author{R.~Soluk}
\affiliation{Physics Department, University of Alberta, Edmonton, Alberta, Canada}

\author{M.~Spurio}
\affiliation{INFN, Section of Bologna \& Department of Physics \& Astronomy, University of Bologna, Italy}

\author{M.~Staelens}
\affiliation{Physics Department, University of Alberta, Edmonton, Alberta, Canada}

\author{M.~Suk}
\affiliation{IEAP, Czech Technical University in Prague, Czech~Republic}

\author{M.~Tenti}
\affiliation{INFN, CNAF, Bologna, Italy}

\author{V.~Togo}
\affiliation{INFN, Section of Bologna, Bologna, Italy}

\author{J.~A.~Tuszy\'{n}ski}
\affiliation{Physics Department, University of Alberta, Edmonton, Alberta, Canada}

\author{A.~Upreti}
\affiliation{Department of Physics and Astronomy, University of Alabama, Tuscaloosa, Alabama, USA}

\author{V.~Vento}
\affiliation{IFIC, Universitat de Val\`{e}ncia - CSIC, Valencia, Spain}

\author{O.~Vives}
\affiliation{IFIC, Universitat de Val\`{e}ncia - CSIC, Valencia, Spain}


\collaboration{THE MoEDAL COLLABORATION}
\noaffiliation

\date{\today}

\begin{abstract}
\noindent
A search for highly electrically charged objects (HECOs) and  magnetic monopoles is presented using 2.2 \fb  of $p-p$ collision data taken at a centre of mass energy  (E$_{CM}$) of 8 TeV  by the MoEDAL  detector during LHC's Run-1. The data were collected using MoEDAL's prototype Nuclear Track Detector array and the Trapping Detector array. The results are interpreted in terms of Drell-Yan pair production of stable HECO and monopole pairs with three spin hypotheses (0, 1/2 and 1). The search provides constraints on the direct production of magnetic monopoles carrying one to four Dirac magnetic charges (4$\gd$) and with mass limits ranging from 590 GeV/c$^{2}$ to  1 TeV/c$^{2}$. Additionally, mass limits are placed on HECOs with charge in the range 10$e$ to 180$e$, where $e$ is the charge of an electron, for masses between 30 GeV/c$^{2}$ and 1 TeV/c$^{2}$.
\end{abstract}


\maketitle
\newpage

The quest for highly ionizing particle (HIP) avatars of physics beyond the Standard Model (SM)  has been an active area of investigation at 
accelerator centres for several decades \cite{Fairbairn:2007,Patrizii-Spurio:2015,Pinfold:2009,Acharya:2014,Aad:2011,Aad:2012, Aad:2013,Aad:2016,Acharya:2016, 
Acharya:2017,Acharya:2018,Acharya:2019,Aad:2015,Chatrchyan:2013,Aaboud:2019, ATLAS-13TeV,Mavromatos:2020}. Searches have also been performed in cosmic rays 
and in matter  \cite{Patrizii-Sahnoun:2019, Burdin:2015}. 

Most HIP searches can be divided into two categories: the quest for magnetic monopoles  (MMs) and the 
hunt for highly electrically charged objects (HECOs). Also, according to the Bethe-Bloch formula  \cite{BB}, massive  singly charged  particles traversing matter can be highly ionizing at low velocity, $\beta$ (the particle velocity expressed  as a fraction of the speed of light, $c$). The physics program of the MoEDAL  experiment described  \cite{Acharya:2014} a number of such massive Singly Electrically Charged Object (SECO)  HIP scenarios.  Further  studies have examined MoEDAL's sensitivity to SUSY SECOs \cite{Felea:2020,Acharya1:2020}  and doubly charged \cite{Acharya1:2020} SUSY particles.

In 1931 Dirac formulated a consistent description  of a magnetic monopole \cite{Dirac:1931} within the framework of quantum physics.  This monopole is  associated with a line of singularity called a Dirac string.  Dirac derived his Quantization Condition (DQC) in order that this string  has no physical effect:
\begin{equation} 
g = ng_{D} = \frac{2\pi\hbar}{\mu_{0}e}n~~~ \textnormal{SI units of Ampere-metres}
\end{equation}
where $e$ is the electric charge of the particle probe, $\hbar$ is Planck's constant divided by 2$\pi$, $g_{D}$ is the magnetic charge, $\mu_{0}$ is the permeability of free space and $n$ is an integer. 

The DQC indicates that if magnetic charge exists then the electric charge is quantized in units of $e = 2\pi\hbar/(\mu_{0}g_{D})$. The value of $g_{D}$ is approximately 68.5$e$.  Dirac's theory did not constrain the mass or the spin of the monopole. Further, the Dirac quantization condition  indicates a coupling strength much bigger than one: $\alpha_{m}= \mu_{0}g_{D}^{2}/(4\pi\hbar c) \approx  $ 34. Thus, perturbation theory cannot be applied and  cross-section   calculations  based on perturbation  theory are not physically valid, although useful as a benchmark.

In 1974 't Hooft \cite{thooft:1974} and Polyakov \cite{polyakov:1974} discovered monopole solutions of the non-Abelian Georgi-Glashow model ~\cite{georgi:1972}. This model   has only one  gauge symmetry, $SO(3)$, with a three component Higgs field. The mass of the 't Hooft-Polyakov MM was predicted to be around 100 GeV/c$^{2}$. However, MMs with such a low mass were  ruled out by experiment. Subsequently, Georgi and Glashow combined their electroweak theory with a theoretical description of strong nuclear forces to form  a Grand Unified Theory (GUT) \cite{georgi:1974} using  the single non-Abelian gauge symmetry, $SU(5)$. In this GUT theory the MM would have a mass of  $\sim$10$^{15}$ GeV/c$^{2}$ which is far too heavy to be directly produced at any foreseeable  terrestrial collider. 
     

The SM  has an $SU(2)\times U(1)$ group structure that does not allow a finite-energy monopole. However,   Cho and co-workers have modified its  structure  to admit the possibility of an ``electroweak'' monopole \cite{Cho:1997,Cho:2005} with a magnetic charge of 2$g_{D}$.  Based on this work, Cho, Kim and Yoon (CKY) \cite{CKY:2015} have more recently presented an adaptation of the SM \textendash   including  a  non-minimal  coupling  of  its Higgs field to  the  square  of  its  $U(1)$  gauge  coupling  strength \textendash  that permits the possibility of a finite energy dyon \cite{Schwinger:1969}. 

In another extension of the SM there exists a topologically stable, finite energy magnetic monopole with a mass estimated to lie in range 900 GeV/c$^{2}$ to 3 TeV/c$^{2}$ \cite{hungmm, mavromatosmm}.  This extension retains the same gauge group as the SM but possesses an extended fermion and Higgs sector where right-handed neutrinos are non-sterile.
  
  The question of whether it is possible to  create generalizations of the CKY model that are consistent with the SM was considered by Ellis, Mavromatos and You (EMY) \cite{EMY:2016}. EMY  concluded that there is a possibility that an ``electroweak" monopole, consistent with the current constraints on the SM, may exist and be detectable at the LHC. The existence of a MM is such a theoretically well  predicated and revolutionary possibility that the search for a MM has been carried out as each new energy frontier is broached.
  
 We consider here only those models that admit a  magnetic charge quantized in units of Dirac charge, $g_{D}$, or a multiple of the Dirac charge.  As $g_{D}$ = 68.5$e$,  a  relativistic monopole with a single Dirac charge will  ionize $\sim$4700 times more than a relativistic  proton.  It is thus a prime example of a HIP. 
  
  As mentioned above electrically charged HIPs, or HECOs, have also been hypothesized. Examples of HECOs, include: dyons \cite{Schwinger:1969},
  doubly charged  massive particles \cite{Acharya:2014}; scalars in neutrino-mass models \cite{Hirsch:2021}; aggregates of $ud$- \cite{Holdom:2018} or $s$-quark matter \cite{Farhi84}, $ Q$-balls \cite{Coleman85,Kusenko98} and the remnants of microscopic black-holes \cite{Koch07}. 
  
  \begin{figure}[h]
    \centering
    \includegraphics [width=0.8\linewidth]{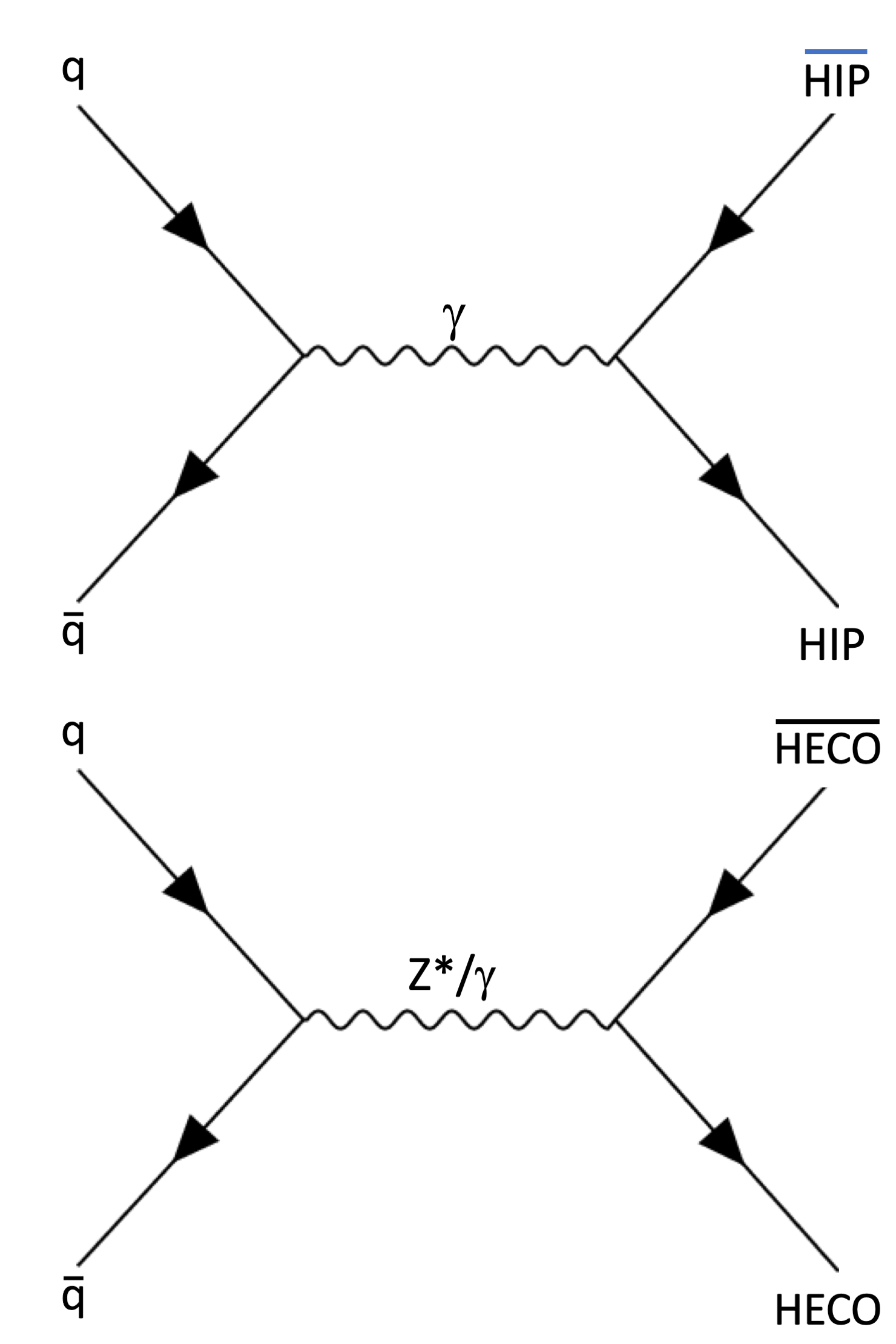}
    \caption{ Tree level Feynman diagram for DY production (top)  of HIP anti-HIP pairs and (bottom) spin-1/2
    HECO pairs.}
    \label{fig:DY-HIPs}
\end{figure}

  The first searches for  MMs and/or HECOs at the LHC were performed by  the ATLAS and MoEDAL  Collaborations in 8 TeV $p-p$ collisions   \cite{Aad:2011,Aad:2012,Acharya:2016}. At this stage, the ATLAS monopole search was sensitive to singly magnetically  charged  (1$g_{D}$)  monopoles,  whereas the MoEDAL search was  sensitive to single and  multiply charged monopoles. ATLAS and MoEDAL  continued the quest for HIPs at Run-2. 
  
  In the case of MMs, the ATLAS and MoEDAL  searches  were complementary, in the sense that ATLAS utilized the MMs  highly ionizing signature \cite{Aad:2016,Aaboud:2019} whereas, until now, the MoEDAL experiment only exploited  the induction technique to directly detect the magnetic charge  \cite{Acharya:2017,Acharya:2018,Acharya:2019}. 
  Extensive accelerator searches  for HIPs at the LHC  have also been undertaken \cite{Aaboud:2019,Aad:2016,Aad:2013,Aad:2011}. The latest result from the LHC is from an ATLAS experiment  search for HECOs and monopoles using data taken during LHC's Run-2 at a centre-of-mass energy of 13 TeV \cite{ATLAS-13TeV}.
  
\begin{figure}[htb]
\begin{center}
 \includegraphics[width=1.0\linewidth]{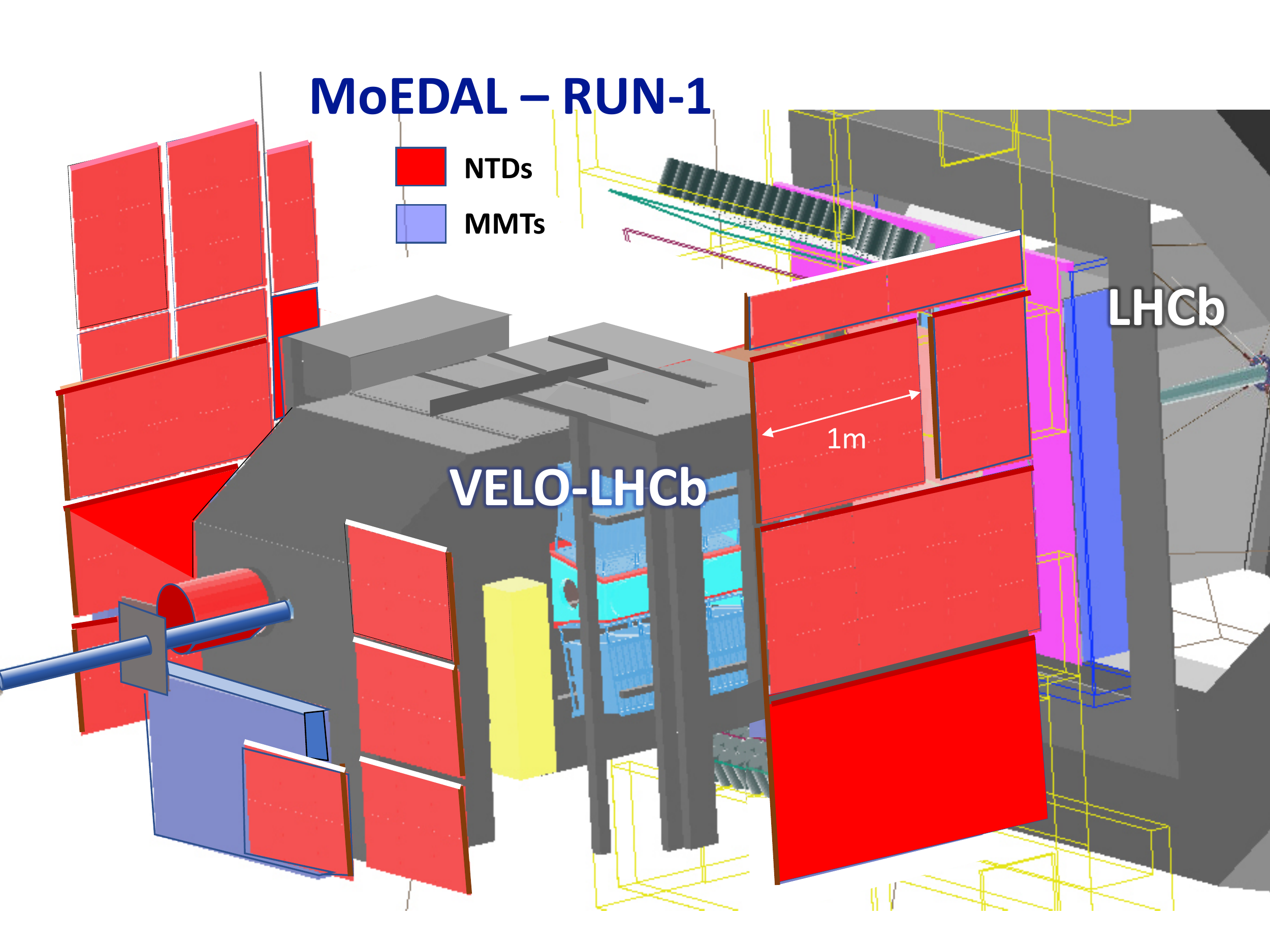} 
 \caption{A GEANT-4 Panoramix view of the  MoEDAL detector prototype deployed at IP8 during LHC's Run-1.}
\label{fig:MoEDAL-3D-views}
\end{center}
\end{figure}

  
   In this paper we report the first use  of the prototype MoEDAL Nuclear Track Detector (NTD)  System, which relies on an ionization signal to detect  HIPs in conjunction with the prototype MoEDAL trapping detector system that utilizes a Superconducting Quantum Interference Device (SQUID)  to detect the  presence of trapped magnetic charge.  The complete prototype detector is shown in Fig.~\ref{fig:MoEDAL-3D-views}. A total of  2.2 fb$^{-1}$ of $p-p$ collision  data  was obtained  during LHC's Run-1  at  intersection point  IP8 on the LHC ring using this detector and analyzed for  evidence of HECOs. The precision of the luminosity measurement  at IP8 during Run-1 is estimated to be 1.16\% \cite{lumi-precision}.
 
   A  DY  mechanism  provides a simple model for HIP pair production. Monopole pair production and spin-0 and spin-1 HECO pair production cross sections are computed using the Feynman-like diagram  shown  in  Fig.~\ref{fig:DY-HIPs}(top). In the case of spin-1/2 HECOs DY production can take place via virtual photon or Z exchange \cite{Wendy}, as depicted in Fig.~\ref{fig:DY-HIPs}(bottom). In the  case of Drell-Yan processes of magnetic monopole production, the coupling of the magnetic charge to the Z boson is usually assumed to be  absent. In specific models of such monopoles this is proven explicitly~\cite{hungmm,mavromatosmm}, since any Z-flux that could exist in the monopole solution would be concentrated inside the monopole core. In the case of electrically charged dyons this issue is model dependent.
   The spin 1 magnetic monopole \cite{Kurochkin-2006,Kurochkin-2007} can have a non-zero magnetic moment, characterized by the parameter $\kappa$ \cite{Baines-2018}. In this analysis the value $\kappa$=1 is used as it is the only one that respects unitarity \cite{Lee-1962}.

   It should be noted that  the large  monopole-photon  coupling   places  such  calculations  in  the non-perturbative regime. In  the case of HECOs, which are characterized by large electrical charges, the Drell-Yan (DY) diagram shown in Fig.~\ref{fig:DY-HIPs} should undergo appropriate resummation, to account for potential non perturbative quantum corrections, see e.g. \cite{Roberts:1994dr}, \cite{Binosi:2009qm}. Such techniques are beyond the scope of this paper, and will be the topic of a future investigation.

\section {The MoEDAL Detector}
MoEDAL's  detector technology is radically different from the general-purpose LHC experiments, ATLAS and CMS. The MoEDAL detector, deployed alongside LHCb's VELO (VErtex LOcator) detector  at IP8, employs two unconventional passive detection methodologies tuned to the discovery of  HIPs. The first of these is a plastic NTD stack array to detect the ionization trail of HIPs. The second is a detector  system comprised  of aluminium absorber elements.  This detector system is called  the MMT (Magnetic Monopole Trapper) since it was used to trap HIPs with magnetic charge, that  slow down and stop within its sensitive volume, for further laboratory analysis. Both of these detector systems are passive, requiring neither  a trigger or readout electronics. The MoEDAL  detector  is described in more detail below. 

The MoEDAL detector is  exemplified by its ability to retain a permanent record, and even capture new particles for further study.
The NTDs provide a tried-and-tested and cost effective method to accurately measure the track of a HIP and its effective charge. Importantly, the NTD response was  directly calibrated using  heavy-ion beams  at the CERN SPS. The second detector system, the MMT, ensures that a small but significant fraction of the HIPs produced are slowed down, stopped and  trapped for further study in the laboratory.  There are no SM particles that can produce such  distinct signatures.  Thus, even the detection in MoEDAL of few HIP  messengers of new physics would herald a discovery. 

\subsection{Energy Loss of HIPs in MoEDAL}
\begin{figure}[htb]
  \centering
    \includegraphics [width=1.0\linewidth]{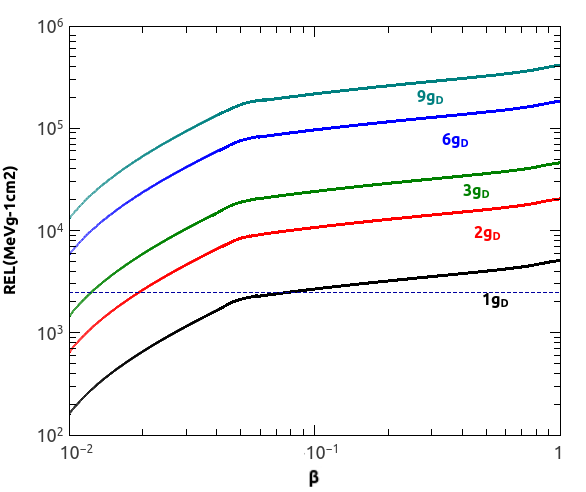}
    \caption{Restricted Energy Loss in  Makrofol for monopoles of different magnetic charge. The horizontal dashed line indicates the Makrofol detection threshold.}
       \label{fig:mm-rel}
       \end{figure}

\begin{figure}[h]
  \centering
    \includegraphics [width=1.0\linewidth]{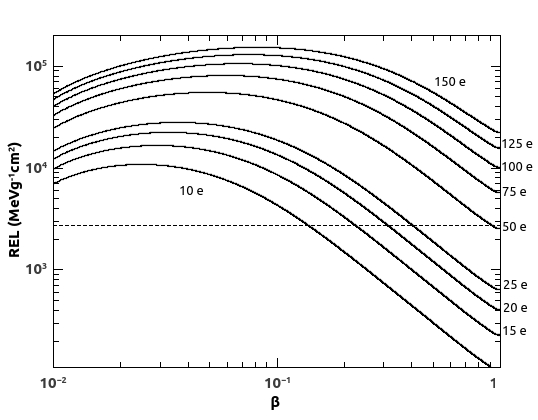}
    \caption{Restricted Energy Loss in Makrofol  for HECOs  of different electric charge. The horizontal dashed line indicates the Makrofol detection threshold.}
       \label{fig:hecos-rel}
       \end{figure}

In the MoEDAL detector HIPs lose energy by ionization. The energy loss by ionization in the MMT detector is computed using Bethe-Bloch formula. For NTDs, the relevant quantity  is the Restricted Energy Loss (REL) \cite{REL}.
For $\beta<$10$^{-2}$, the REL is equal to the particle's total energy loss  in the medium.  At larger velocities, REL is the  fraction of the electronic energy loss leading to the formation of $\delta$-rays with energies lower than a cut-off energy T$_{cut}$. The
REL can be computed from the Bethe-Bloch formula restricted to energy transfers T$<$T$_{cut}$ with T$_{cut}$ a constant characteristic of the medium. For  Makfrofol, which is the MoEDAL NTD used for the analysis reported in this paper, T$_{cut} \le$ 350 eV.
The RELs for MMs and for HECOs in Makrofol   are shown in Fig.~\ref{fig:mm-rel} and Fig.~\ref{fig:hecos-rel}, respectively.

\subsection{The MMT Detector}
\begin{figure}[h]
  \centering
    \includegraphics [width=1.0\linewidth]{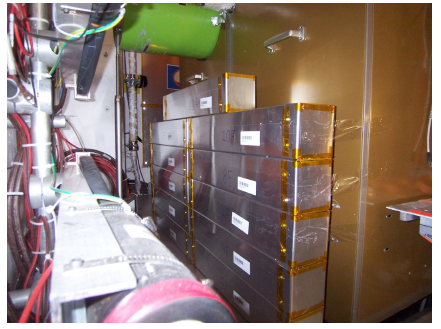}
    \caption{A photograph of the prototype MMT detector deployed at IP8.}
       \label{fig:MMTprototype}
       \end{figure}

The prototype MMT detector deployed for LHC's Run-1 consisted of 198 aluminium rods weighing a total of 163 kg. These rods were housed in an enclosure placed just underneath the beampipe at the upstream end of LHCb's VELO detector as shown in  Fig.~\ref{fig:MMTprototype}.
After exposure the MMT's aluminium volumes are sent to the ETH Zurich Laboratory for Natural Magnetism where they are passed through a SQUID magnetometer to scan  for the presence of trapped magnetic charge. A monopole will stop in the MMT detector  when  its speed falls below  $\beta \le$ 10$^{-3}$. It then binds due to the interaction between the monopole and the nuclear magnetic moment\cite{binding,goebel:1984,bracci:1984,olaussen:1985} of an aluminium nucleus comprising a MMT trapping volume. 

The anomalously large  magnetic moment of an aluminium nucleus gives rise to a monopole-nucleus binding energy  of 0.5 -  2.5 MeV \cite{binding},  comparable to the shell model splittings. In any case, it is reasonable to assume that the very strong magnetic field  of the monopole will rearrange the nucleus, permitting it to bind strongly to the nucleus. As reported in  Ref. \cite{binding} monopoles bound in such a way would  be trapped indefinitely. It would require  fields well in excess of several Tesla for the lifetime of the   trapped monopole state to compromise its  detection by the MoEDAL trapping detector. We note that the MoEDAL detector is  only subject to fields lower than $\sim$10 mT.

\subsubsection{Calibration of the MMT Detector}
A magnetic monopole captured  in an MMT volume is tagged and  measured as a persistent current in the SQUID coil encircling the samples' transport axis that  passes through the SQUID magnetometer.
The calibration of the magnetometer response is achieved using two independent techniques. In brief, the magnetometer calibration was obtained using a convolution method applied to a dipole sample, and validated using long thin solenoids that simulate a monopole of well-known magnetic charge. For more details see Ref.~\cite{DeRoeck}.
These calibration  methods agree to within 10\%, which is taken  as the pole strength  calibration uncertainty.  The magnetometer response has been determined by measurement to be  charge-symmetric and linear in a range of magnetic charge 0.3 - 300 $\gd$.









\subsection{The  Nuclear Track Detector System}
\label{NTD}

\begin{figure}[h]
    \centering
    \includegraphics [width=1.0\linewidth]{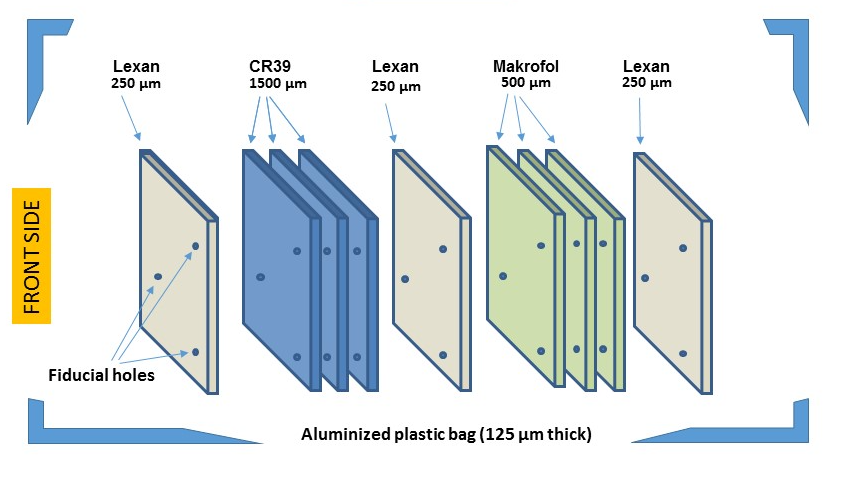}
    \caption{ NTD module composition}
    \label{fig:ntd-unit}
\end{figure}

 The MoEDAL NTD is arranged in modules
deployed around IP8 in the VELO  cavern. A  prototype NTD array of 125 $\times$ 25 cm $\times$ 25 cm stacks was installed for Run-1 as shown in  Fig.~\ref{fig:MoEDAL-3D-views}.
 Each module comprises three layers of 1.5 mm thick CR39\textsuperscript{\textregistered} polymer,   three layers of Makrofol DE\textsuperscript{\textregistered} and three layers of Lexan\textsuperscript{\textregistered} 0.5 and 0.25 mm thick, respectively, inside Aluminium bags (Fig.~\ref{fig:ntd-unit}).  Currently the Lexan\textsuperscript{\textregistered}  foils serve as protective layers and are not analyzed.
 
 In this analysis only the Makrofol NTDs are utilized. This is due to roughly a factor ten higher detection threshold in Makrofol than CR39 which results in substantially less ``visual noise'' in the etched plastic large  due to  spallation products arising from beam backgrounds. Beam background particles are generated by the interaction of the LHC beam with LHC machine elements and the machine  environment. Thus, the analysis of the CR39 NTDs is
 considerably more time intensive. HIPs produced via the DY mechanism 
 in LHC collisions during Run-1 are sufficiently highly ionizing that they can easily be detected with the  Makrofol NTDs, obviating the need to scan the CR39 in the
 first pass. In the event of the observation of a candidate event in the Makrofol all 6 NTD sheets in the stack would have been analyzed.

 
\par
\subsubsection{The Etching Procedure}
\begin{figure}[h]
    \centering
    \includegraphics [width=1.0\linewidth]{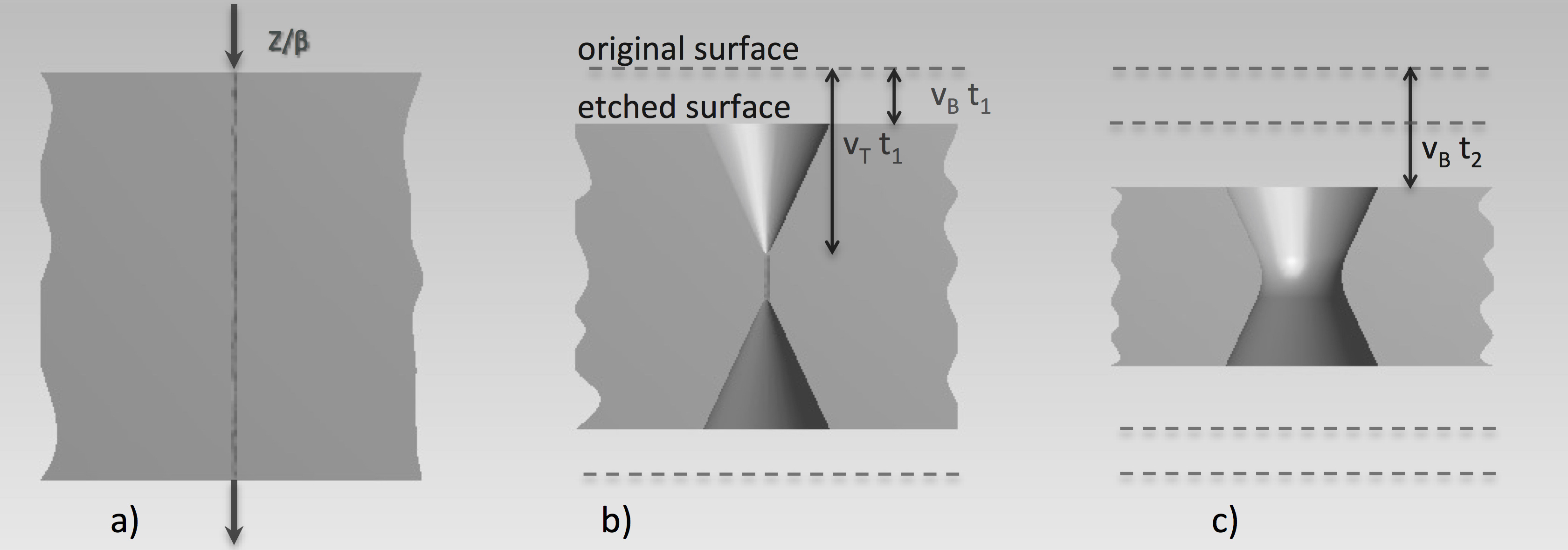}
    \caption {Illustration of the track-etch technique: a) latent track forming along the trajectory of a high ionizing  particle impinging perpendicularly on the NTD surface ; b) development of conical pits during the etching process; c) etch-pits joining after a prolonged etching, forming a hole in the detector. }
   \label{fig:ntd-method}
\end{figure}

In plastic track-etch detectors, the passage of a heavily ionizing particle can produce a permanent damage of polymeric bonds in  a cylindrical region (``latent track") extending few tens of nanometers around the particle trajectory (Fig.~\ref{fig:ntd-method}). By subsequent chemical etching 
the latent track is ``amplified" and can be made visible under an optical microscope. In the etching process,
the bulk of the material is removed at a rate $v_B$ and at a higher rate $v_T$ along the latent track. The damage zone is revealed under an optical microscope as  a pair of cone shaped etch-pits, one on each face of the NTD sheet.  Etch-pits surface openings have a circular shape for normally incident particles, otherwise they are elliptical. A single well measured etch pit is called a ``track'' candidate. If another etch pit is measured on the NTD  sheet that is consistent with being the twin then we have a confirmed track candidate.

\begin{figure}[h]
    \centering
    \includegraphics [width=1.0\linewidth]{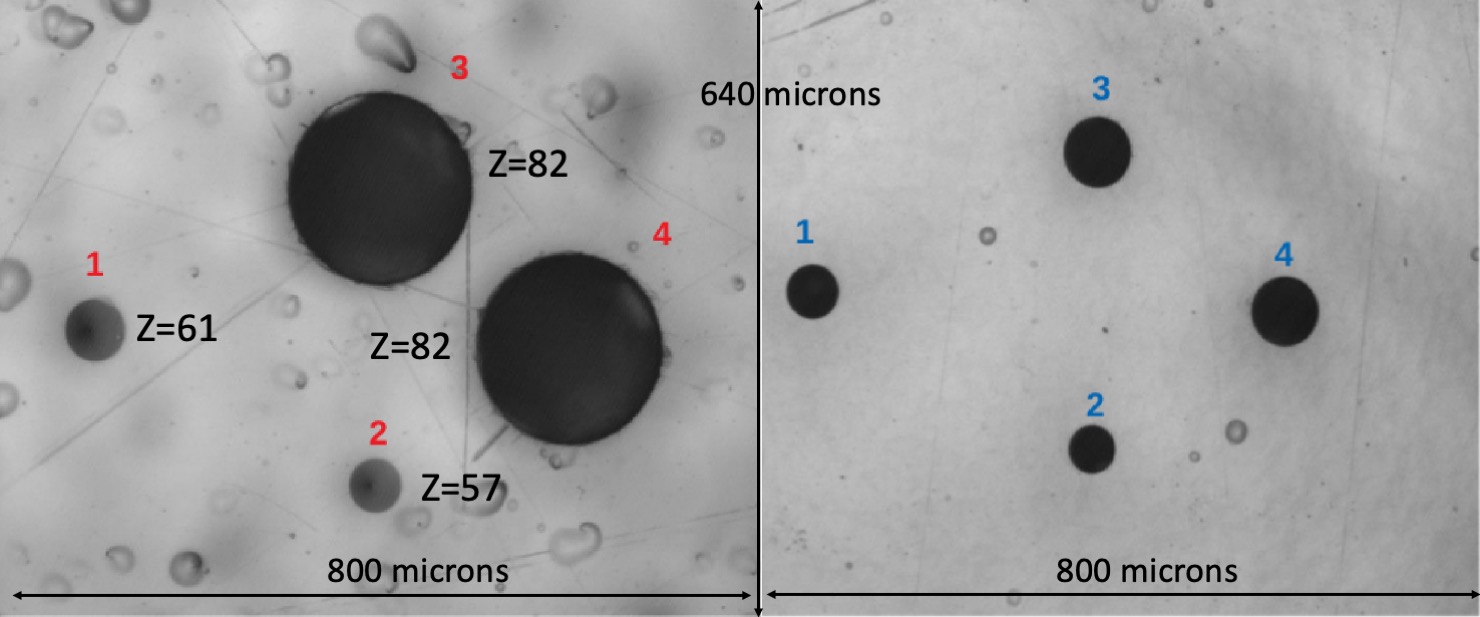}
    \caption{Microphotographs of relativistic Pb$^{82+}$  tracks and of nuclear fragments (Z$<$82) in  two consecutive foils of Makrofol. Each image frame measures 0.64 mm x 0.80 mm. Etch-pits are from the same ions crossing the detector foils: 
    (left) Makrofol foil etched  in ``strong conditions''; (right) Makrofol foil etched in ``soft conditions'' . Note that the microphotographs also  show two clearly differentiated  fragmentation products of Pb:  La (Z = 57); and,  Pm (Z=61). }
    \label{fig:etch-pit-photo}
\end{figure}

 A sketch of an etch-pit at different etching times is shown in Fig.~\ref{fig:ntd-method} for a normally incident particle crossing the detector with a constant energy loss. Two etching conditions were applied (Table \ref{tab:etching-conditions}).  The first is the so-called ``strong'' etching condition,  allowing faster etching and yielding larger etch-pits that were easier to detect under visual scanning. Strong etching  was applied to the first, most upstream,  Makrofol foil in each module. 
 The second, ``soft etching,''  condition results in a slower etching process. This allows the etching process to proceed in several steps in order  to follow the formation of etch-pits. Soft etching is applied to subsequent Makrofol foils in the stack, if a candidate track is found in the first layer.  In Fig.~\ref{fig:etch-pit-photo} are shown microphotographs of relativistic Pb$^{82+}$ tracks in Makrofol foils etched in (left)  ``strong conditions''; (right) ``soft conditions.''


\begin{table*}[htb]
\caption{\label{tab:etching-conditions} Etching Conditions of Makrofol}
\centering
\begin{tabular}{|c|c|c|} 
\hline
\multicolumn{3}{|c|}{Etching Conditions} \\
\hline
Etching Mode & Etchant & v$_B$($\mu$m/hour)\\
\hline
Strong  & 6N KOH + 20\% ethyl alcohol at 65$^{\circ}$C & 23$\pm$0.5\\
Soft & 6N KOH + 20\% ethyl alcohol at 50$^{\circ}$C & 3.4$\pm$0.05\\
\hline
\end{tabular}
 \end{table*}

\subsubsection{Calibration of the NTD Detector} \label{subsubsection:calibration}
The response of the NTD  is measured by the etching rate ratio, also called the reduced etch-rate,  $p$ = v$_T$/v$_B$, as a function of the particle's REL. Heavy ion beams are  used to determine the detector response over a large range of energy losses, as discussed in ref. \cite{calib2007}.
The Makrofol  was calibrated with  ion beams  of 158 A GeV Pb$^{82+}$ and 13 A GeV Xe$^{54+}$ energy per nucleon, at CERN's SPS. The calibration set-up included a stack of Makrofol foils placed upstream  and downstream  of an Aluminum target.  Incoming ions undergo charge changing nuclear fragmentation interaction along their path through the detector foils and the target. After exposure the detectors were etched in 6N KOH+20$\%$ ethyl alcohol at 50~$^\circ$C for 10 hours. The bulk etching velocity was v$_B$ = 3.4~$\mu$m/h.

\begin{figure}[htb]
   \centering
    \includegraphics [width=0.9\linewidth]{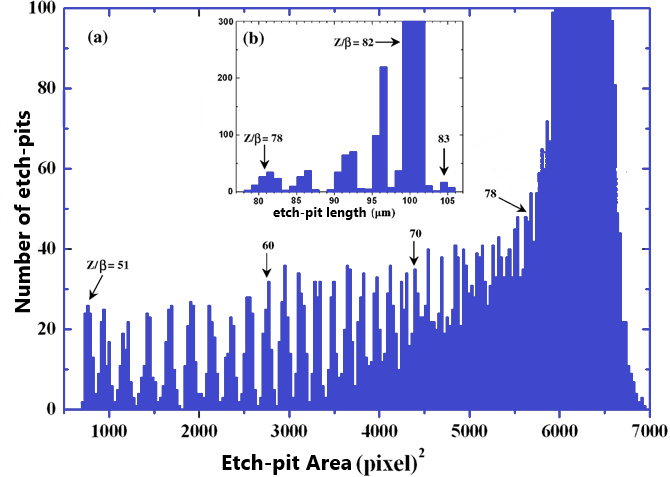}
    \caption{Distribution of track surface areas in Makrofol exposed to 158 A GeV Pb$^{82+}$ and etched in soft conditions \cite{calib2007}.}
   \label{fig:mak-peaks}
   \end{figure}

After etching, the size of the etch pits  was  measured with an automatic scanning system providing the cone base area, and the coordinates of the center of the etch pits.   Etch pits diameters typically  range from 10 $\mu$m to 100 $\mu$m,  with a modal value in the range 30 $\mu$m to 40 $\mu$m. 
The  base area distributions of incoming ions and of their fragments is shown in Fig.~\ref{fig:mak-peaks}.
The projectile fragments have the same velocity and approximately the same direction as the incident ions.  From the base area spectrum, the  charge corresponding to each  nuclear fragment peak can be identified, and the corresponding REL determined. 
A detailed description of the calibration procedure can be found in \cite{calib2007}.

\vskip 1cm

\begin{figure}[h]
\includegraphics[width=0.9\linewidth]{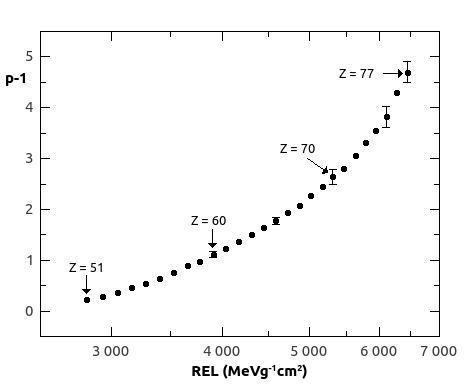}
\includegraphics[width=0.9\linewidth]{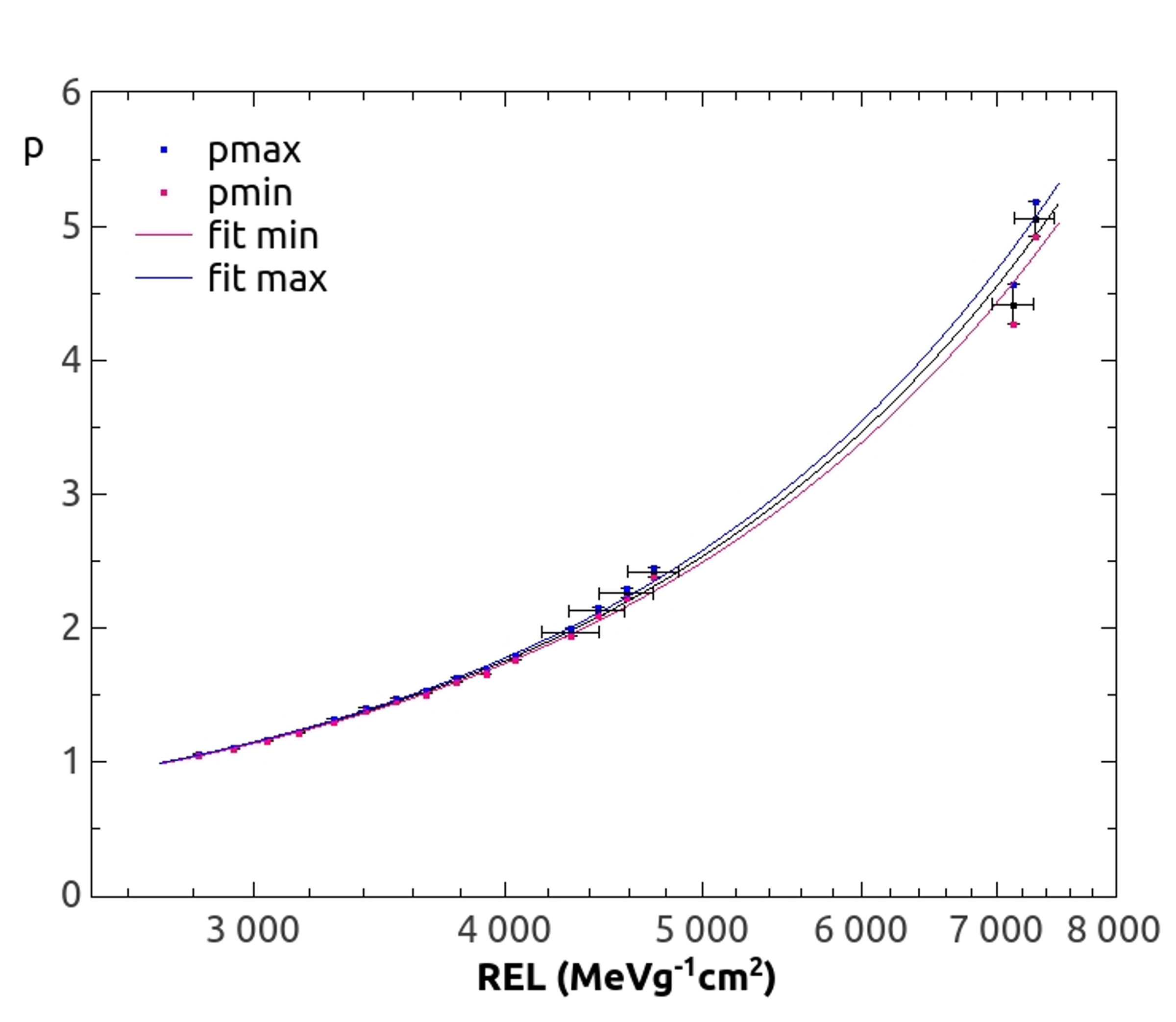}
\caption {\label{fig:mak-calibs}Reduced etch-rate ($p$) versus REL for Makrofol exposed to relativistic Lead and Xenon ion beams: (top) detectors etched in soft conditions;(bottom) detectors etched in strong conditions. The upper and low curves are drawn through the $\pm$1$\sigma$ value of the  error on each  $p$ value, where the error bars  represent  a convolution of the statistical  and systematic errors on each point.} 
\end{figure}

For each identified peak the reduced etch rate p, the Z/$\beta$ and eventually the restricted energy loss are computed (Fig.\ref{fig:mak-peaks}).
Calibration data  thus obtained are shown  in Fig.\ref{fig:mak-calibs}. For Makrofol, the mimimum detectable relativistic charge is Z/$\beta \sim$ 50, for both soft and  strong  etching.  The REL corresponding to this detector threshold  is ~$\sim$ 2700~MeV~g$^{-1}$cm$^{2}$.


\subsubsection{Etching and Scanning of MoEDAL NTD}
After exposure in the LHC IP8 region, the MoEDAL NTD stacks were taken  to the INFN etching and scanning laboratory in Bologna.  A global module reference system  is created by drilling three reference holes  -- 2 mm diameter --  on each detector module. This coordinate system provides an accuracy of 100~$\mu$m   on the determination of the position of a particle track over the detector surface. The stacks are then unpacked, the detectors foils labelled and their thickness measured on a grid of points uniformly distributed over the foil surface.\par
 
For the search reported in this paper  only  Makrofol foils were analysed.  In each exposed stack, the  most upstream Makrofol layer  was etched 
in 6 N KOH + 20\% ethyl alcohol at 65$^{\circ}$C. After 6 hours etching, etch-pits as small as 10$\mu$m would be detected under 20~$\times$ magnification. An efficiency of $\sim$ 99\% was estimated by scanning foils exposed to ions as described below.
 
Each Makrofol layer examined was manually scanned.  Every detected surface structure   was further observed under higher magnification  and classified  either as material defects or particle's track. If a pair of etch-pits is detected on the front and back sides of the foil, it was  observed at larger (100 - 500$\times$) magnification.
From the etch-pit size 
on the ``front'' and ``back'' surfaces,  and the bulk etching rate, the incidence angle on each surface is estimated. It takes around 2.5 hours to scan one 
side of a Makrofol layer  when using the 
microscope at 32$\times$ total magnification.

A pair  of collinear  incident  and exiting etch-pits, consistent with pointing to the IP,  is defined as a potential 
candidate ``track''. In particular, if a candidate is  found in the first layer of a module the  downstream Makrofol foils would  be  etched in 6 N KOH + 20\% ethyl alcohol at 50$^{\circ}$C. The vast majority of spallation products arising from beam backgrounds have a very limited range  in the NTD sheet, typically tens of microns, and only give rise to a single pit when the NTD sheet is etched.

  An accurate scan of the downstream  Makrofol sheets would then be performed using an optical microscope with high magnification (100 - 200~$\times$)  in a square region of about 1 cm$^2$ around the candidate expected position.   If collinear etch-pits are found in all three Makrofol sheets the CR39 would then be scanned for etch-pits collinear with those in the  Makrofol layers. A HIP candidate track requires collinear etch-pits in all six NTD sheets in the stack that points to the IP. Additionally, the REL estimated from the etch-pits dimensions (surface area, etch-pit length)  has to be consistent with the HIP hypothesis. However, no  candidate ``track''  candidate was found.
  
 

 \subsubsection{The Detection Threshold for Makrofol}
\begin{figure}[htb]
   \centering
    \includegraphics [width=0.95\linewidth]{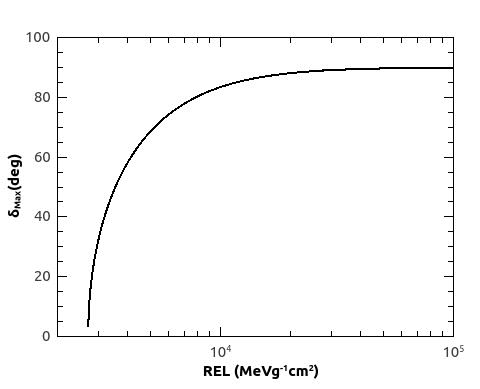}
   \caption{The maximum angle to the normal of the NTD plane within which the HIP will be detected.}
       \label{fig:REL-curve}
   \end{figure}

 For the HIP to be detected its REL must be greater than the detection threshold of the Makrofol. The detection threshold
  will vary with the etching conditions.   It will also  vary with the angle of incidence ($\delta$) of the HIP  on the NTD. The 
  connection between the threshold and the maximum angle of incidence ($\delta_{Max}$) to the normal to the NTD that
   the HIP can make and still be detected, is expressed by the relationship: $ p = \frac{1}{\cos(\delta_{Max})}$, where $p$ is 
  reduced etch-rate  described above. 
  The lowest threshold is obtained 
    for a HIP impinging normally to the NTD.  The curve obtained from a empirically based  parameterization of  the relation between  $\delta_{Max}$ and the REL is shown
     in Fig.~\ref{fig:REL-curve}. This parameterization is used in the determination of  the acceptance for HIPs incident on the MoEDAL's NTD stacks.
 
\vskip 1cm

\subsubsection{Efficiency and False Positives in the NTD Detectors}
As described above the signal for the passage of a HIP messenger of new physics through a MoEDAL 
NTD stack would be a string of etch-pits  in the stack, where an etch-pit pair is  due to the ingress 
and egress of the HIP passing though an NTD sheet. No such signal has ever been seen in this search,
or observed by any other HIP search employing NTDs \cite{Patrizii-Spurio:2015}. Indeed, no candidates
were  seen  in the 125 stacks examined (corresponding to 7.8 m$^{2}$), where only  the most upstream sheet of the NTD stacks were examined.  

The absence of false positives using the NTD technique was also a feature of the
astroparticle physics experiments MACRO \cite{MACRO} and SLIM \cite{SLIM} , which deployed  a surface area 
of 1263 m$^{2}$  and 427 m$^{2}$, respectively. Neither experiment  observed a single HIP candidate. It should be noted
that the NTD technique employed by these experiments are essentially identical to those employed at colliders.

The lack of false positives in the NTD technique at colliders or in astroparticle physics experiment 
raises the question of the false negatives  or detector efficiency, where a signal exists but is not seen. This can be evaluated using the heavy ion beams that
 are used to calibrate NTD detectors. In the absence of beam backgrounds, the detection, or scanning,  efficiency for the
  etch-pits due to heavy-ion HIPs with ionizing power  above the NTD threshold was  measured to be in excess of 99\% as described directly below. 

In order to estimate the  detection efficiency of NTDs for HIPs  in the presence of beam backgrounds we utilized NTD  calibration 
stacks exposed  to a relativistic  lead-ion beam as described above. The stacks were comprised of sheets of  Makrofol 
NTDs exposed to the beam backgrounds (LHC-exposed sheets)  in  the VELO  cavern at the LHC for a year of data taking, interleaved with $\it{unexposed}$
 Makrofol NTD sheets  (LHC-unexposed sheets).  Plastic from the same production  batch  was used in calibration and standard data taking.
 
 The  NTDs sheets comprising the calibration stacks  were  then  etched  in the same way as the  standard NTD stacks 
 deployed for data taking during Run-1. The individual sheets  were  scanned using the same  manually controlled optical scanning 
 microscope technology employed to examine  all MoEDAL NTD stacks. 
 
 The relativistic lead-ion calibration  beam particles  penetrate 
 the whole stack allowing the signal etch-pits  seen in the LHC unexposed sheets   - where the signal
  can clearly be observed  with a 100\% efficiency -  to serve as a map.  The identification of etch-pits in the LHC-unexposed 
  sheets is measured  to be 100\% by making independent comparison scans of the  other LHC-unexposed sheets in the stack which, of course,
  have the  identical etch-pit number and pattern.
  
  Using the LHC-unexposed sheets in the stack as a one-to-one same-scale map for the hits in the adjacent LHC-exposed sheets the scanning efficiency for 
  LHC-exposed sheets can be measured. Such measurements indicate that the overall scanning  efficiency for detection above threshold was in excess of 99\%. 
  This number was found by scanning the LHC-exposed sheets and then comparing the etch-pits found  with the etch-pits identified  in the adjacent 
  LHC- unexposed stacks.  Each sheet has exactly the same number and pattern of `` signal'' etch-pits, resulting from the calibration beam  - since 
 the beam  passes through the complete stack. 

\section {Acceptance of the Run-1 MoEDAL Detector}
The MoEDAL detector's acceptance  is defined to be the fraction of the number of events in which at least one HIP of the
 DY produced  pair was detected in MoEDAL in either the NTD detector or the MMT detector.  The 
 acceptance for DY production of HECOs and magnetic monopoles is described  by an interplay of the geometrical disposition 
 of MoEDAL NTD modules and MMT detectors, energy loss in the detectors, mass of the 
 particle and the spin-dependent kinematics of the interaction products. In the case of the HECOs, MoEDAL's NTD system  provides 
 the only means of detection. 

\begin{figure}[ht]
\begin{center}
 \includegraphics[width=1.0\linewidth]{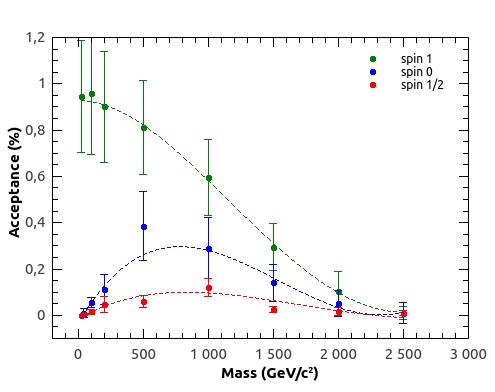}
 \caption{Acceptance for  spin-1, spin-0 and spin-1/2 HECOs  with charge 125$e$, produced via a DY process with virtual photon exchange only. The dashed  lines represent 4th order polynomial fits to the data.} 
\label{fig:heco-acceptance}
\end{center}
\end{figure}

\begin{figure}[htb]
\begin{center}
 \includegraphics[width=1.0\linewidth]{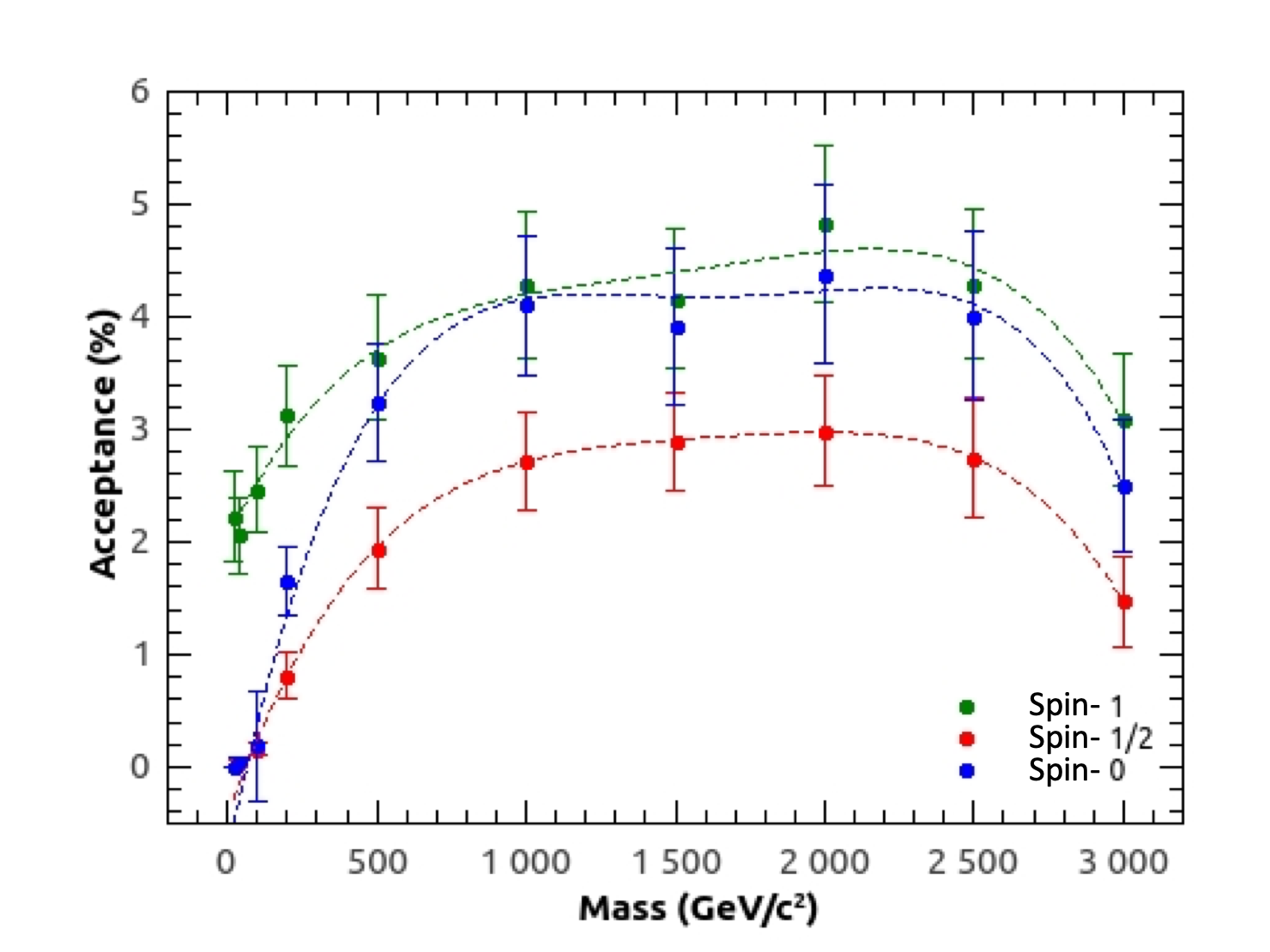}
 \caption{Acceptance for monopole pair production with magnetic charge 2$\gd$. The dashed lines represent 4th order polynomial fits to the data.}
\label{fig:monopole-acceptance}
\end{center}
\end{figure}

For a given HIP  mass and charge, the pair-production model determines the kinematics and the overall trapping acceptance obtained. The uncertainty in the acceptance is dominated by uncertainties in the material description~\cite{Acharya:2016,Acharya:2017,Acharya:2018}. This contribution is estimated by performing simulations with hypothetical material conservatively added and removed from the nominal geometry model. An example, showing the MoEDAL NTD acceptance curves for spin-1/2, spin-0, spin-1 HECOs with charge 125$e$ - produced by a DY process via virtual photon exchange -  is shown in Fig.~\ref{fig:heco-acceptance}. Note that HECOs can only be detected by the NTDs since  HECOs trapped in the MMT detector do not have a magnetic charge and hence cannot be detected in MoEDAL's SQUID detector.  These curves are  determined by the kinematics of the produced particles convoluted with the VELO material immediately surrounding IP8 and the  distribution of the MoEDAL NTD detectors.  The prototype MoEDAL detector provides only a partial, non-uniform,  coverage of the available solid angle.  
 
\begin{figure}[htb]
\begin{center}
 \includegraphics[width=1.0\linewidth]{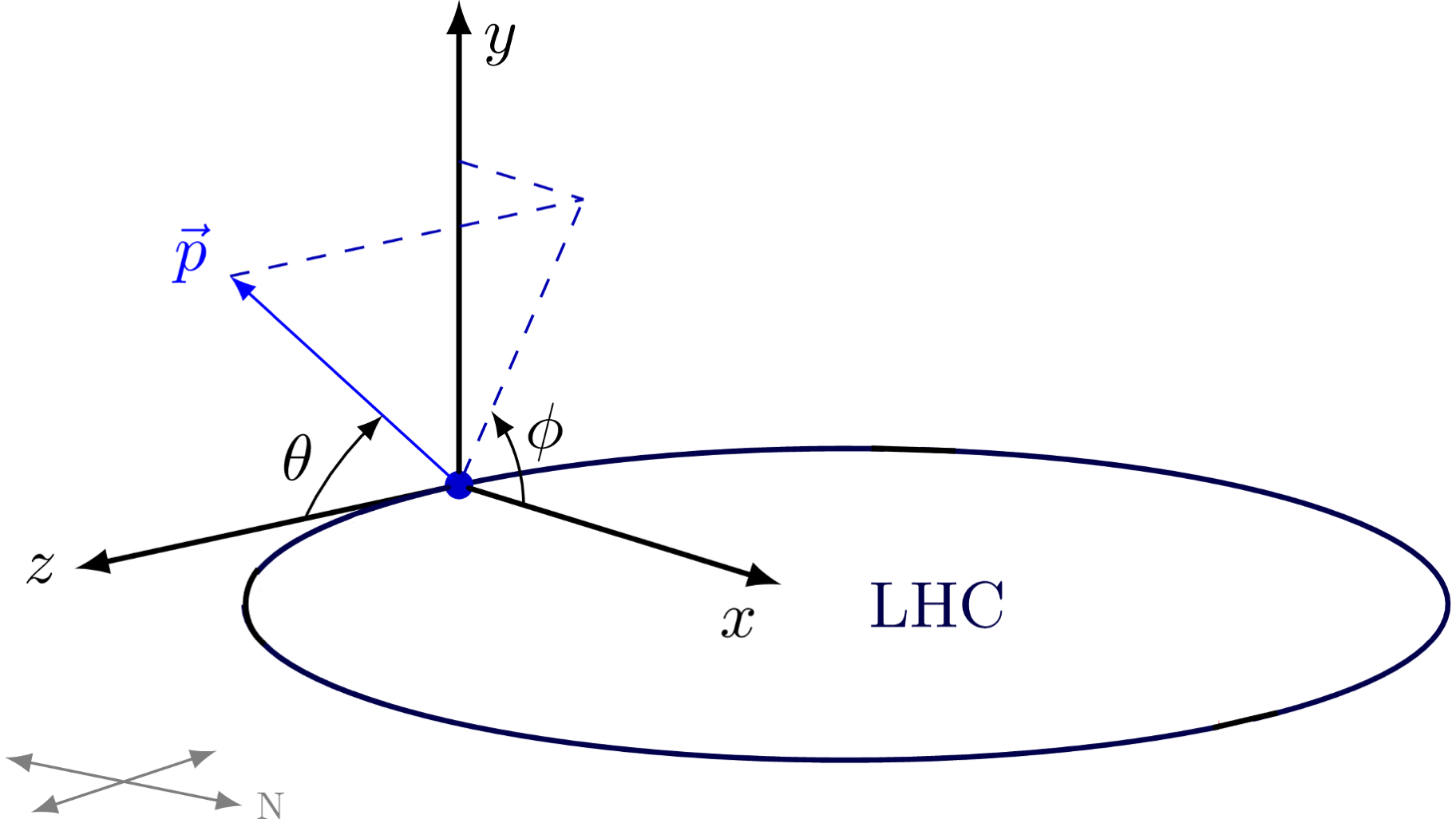}
  \caption{The coordinate system used in the analysis}
\label{fig:coords}
\end{center}
\end{figure}

The acceptance curves for spin-0, spin-1/2 and spin-1 monopoles, found using the NTD and MMT detectors combined, are shown in Fig.~\ref{fig:monopole-acceptance}.  
The acceptances shown in Fig.~\ref{fig:heco-acceptance} and  Fig.~\ref{fig:monopole-acceptance} refer to the prototype detector deployed for LHC's Run-1. The acceptance for the Full Run-2 detector is somewhat larger. The difference between the acceptances for HECOs with  spins with spin-1/2, spin-0 and spin-1, is mainly due to how well the disposition of the detector elements in theta ($\theta$) matches the corresponding theta distribution of the DY produced HECO's. To illustrate this point  we have compared  the kinematic quantities (momentum ($\overrightarrow{p}$), theta ($\theta$) and phi ($\phi$) ) of the uncut signal to the corresponding distributions for the events that pass the selection criteria.  The angles theta ($\theta$) and phi ($\phi$)  are defined in Figure~\ref{fig:coords}. 

\begin{figure*}[htb]
\begin{center}
 \includegraphics[width=1.0\linewidth]{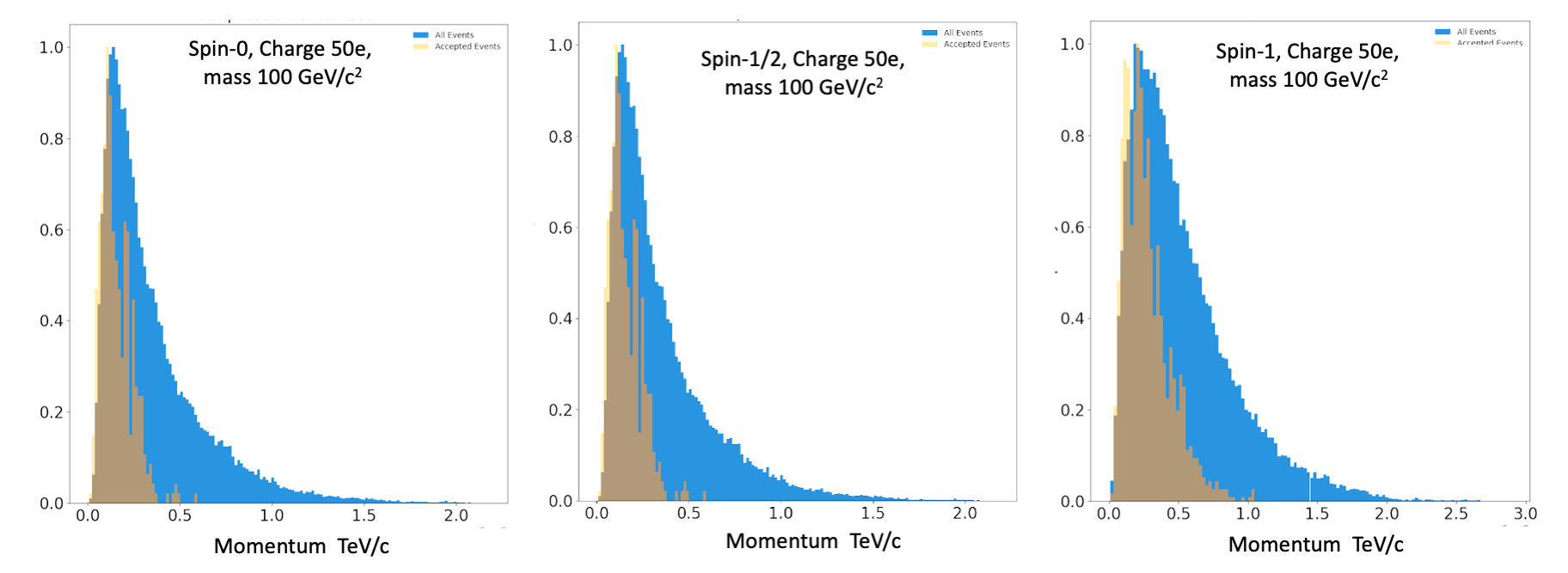}
 \caption{The momentum distribution for  HECOs of mass  100 GeV/c$^{2}$  produced via the DY process with virtual photon exchange, with electric charge $50e$,  for each spin assignment. In each case,  60K events were originally generated. The plots are normalized to the maximum amplitude. The blue histogram represents generated events and the underlying light brown histogram shows the distribution of selected events.}
\label{fig:HECOP}
\end{center}
\end{figure*}

\begin{figure*}[htb]
\begin{center}
 \includegraphics[width=1.0\linewidth]{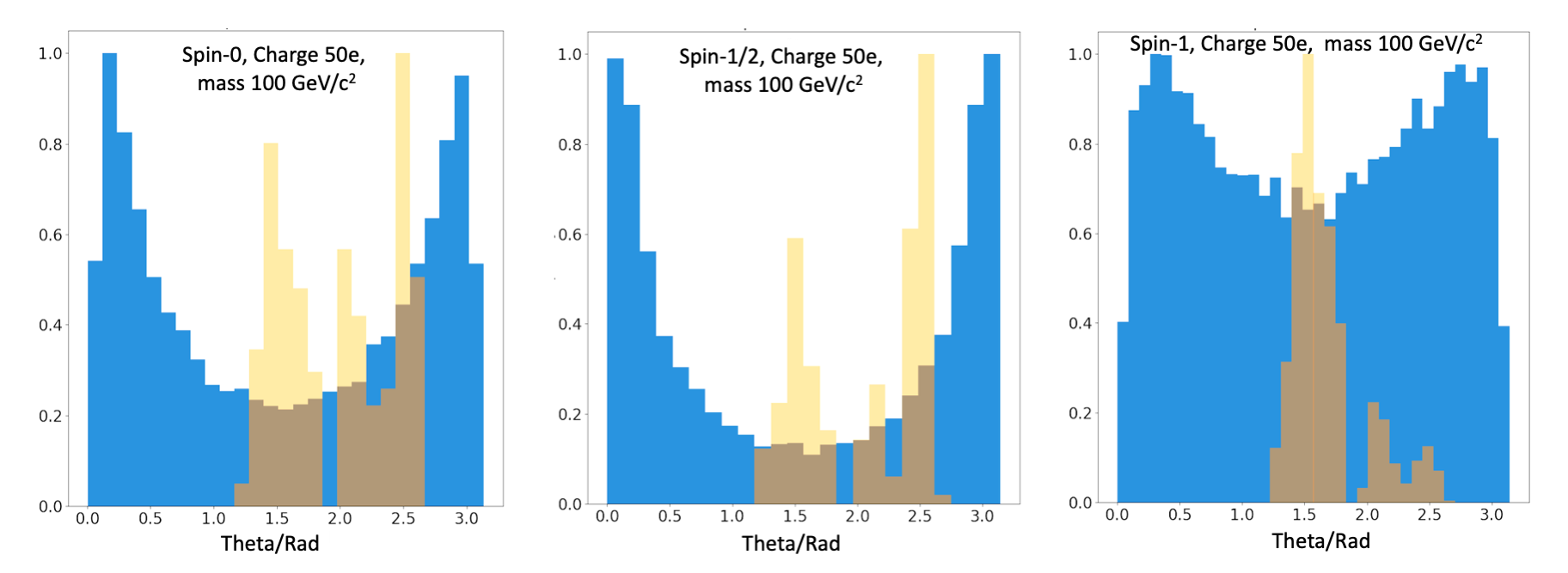}
 \caption{The theta angle distribution for  HECOs of mass  100 GeV/c$^{2}$  with electric charge 50$e$  produced via the DY process vias virtual photon exchange,  for each spin assignment.  In each case,  60K events were originally generated. The plots are normalized to the maximum amplitude. The blue histogram represents generated events and the underlying light brown histogram shows the distribution of selected events. }
\label{fig:HECOTHETA}
\end{center}
\end{figure*}

\begin{figure*}[htb]
\begin{center}
 \includegraphics[width=1.0\linewidth]{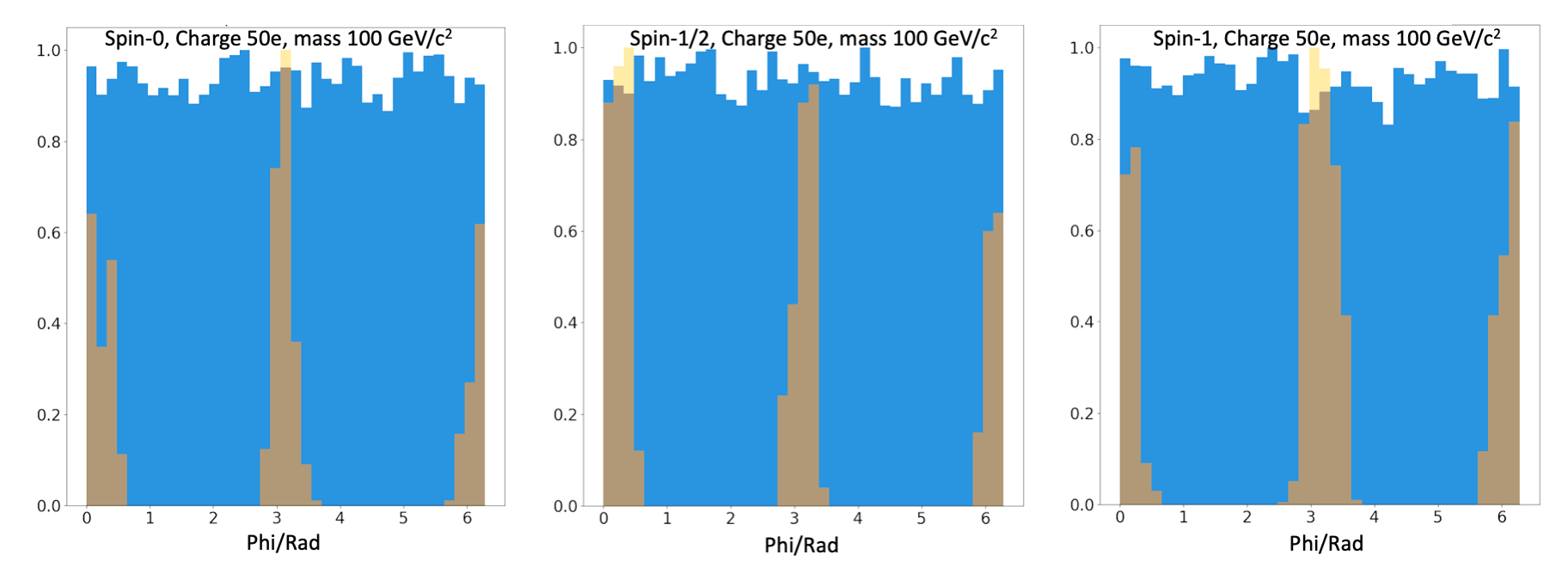}
 \caption{The phi angle distribution for  HECOs of mass  100 GeV/c$^{2}$   with electric charge 50$e$,  produced via the DY process with virtual photon exchange,  for each spin assignment.  In each case,  60K events were originally generated. The plots are normalized to the maximum amplitude. The blue histogram represents generated events and the underlying light brown histogram shows the distribution of selected events. }
\label{fig:HECOPHI}
\end{center}
\end{figure*}

\begin{figure*}[htb]
\begin{center}
 \includegraphics[width=1.0\linewidth]{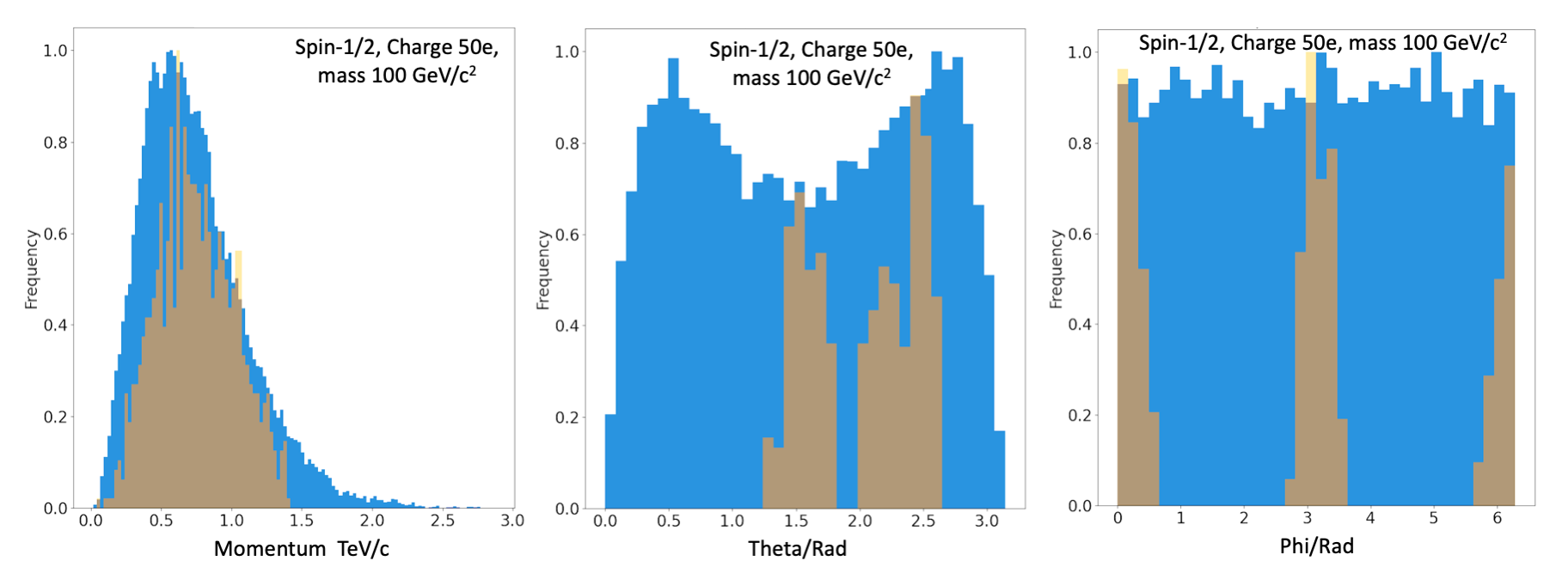}
 \caption{The momentum, theta and  phi angle distribution for  spin-1/2 HECOs of mass  100 GeV/c$^{2}$  and  electric charge 50$e$,  produced via the DY process via virtual photon and Z$^{0}$ exchange.  In each case,  60K events were originally generated. The plots are normalized to the maximum amplitude. The blue histogram represents generated events and the underlying light brown histogram shows the distribution of selected events. }
\label{fig:HECOGZ}
\end{center}
\end{figure*}

In order to make the comparison clearer, given the large disparity between the number of generated and selected events, each plot was normalized to the the highest amplitude bins as selected by the Freedman - Diaconis rule \cite{Freedman-Diaconis}. As an example we plotted the kinematic variables for DY production of  HECOs  with   mass of 100 GeV/c$^{2}$ and electric charge 50$e$   are  shown  in Figure~\ref{fig:HECOP}, Figure~\ref{fig:HECOTHETA} and Figure~\ref{fig:HECOPHI}. 
The corresponding plots for  DY production via  photon/Z$^{0}$* exchange, which affects only the spin-1/2 case, are shown in  Figure~\ref{fig:HECOGZ}.

\begin{figure*}[!htb]
  \begin{center}
  \includegraphics[width=0.47\linewidth]{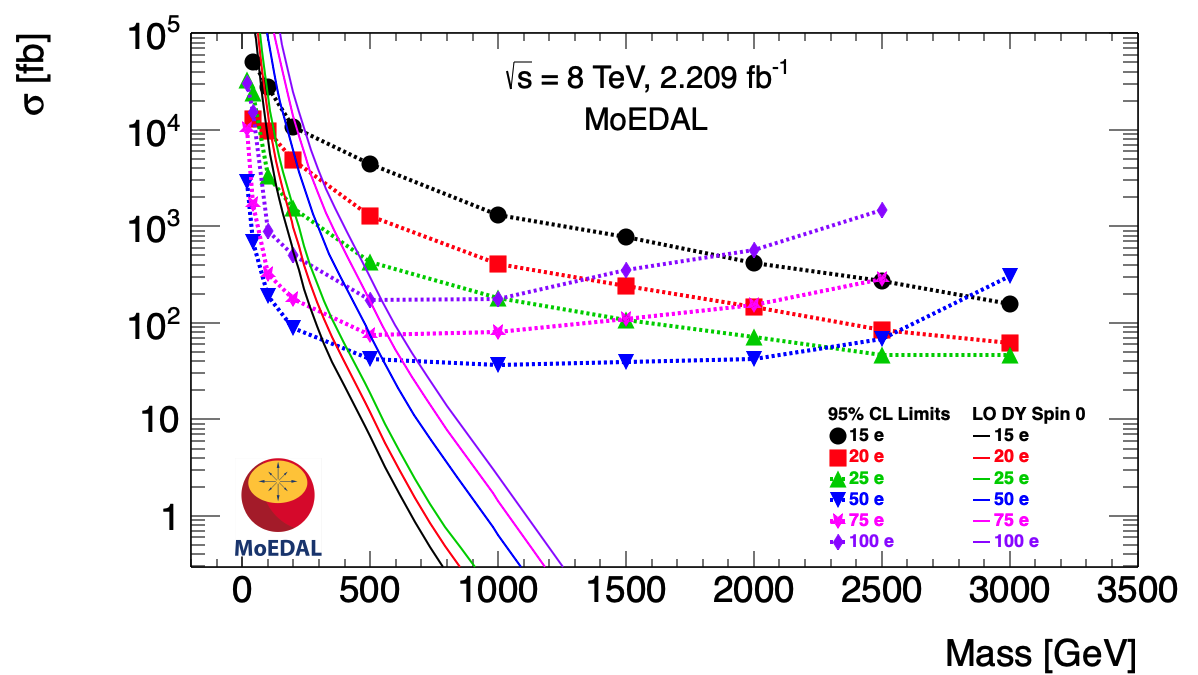}  
    \includegraphics[width=0.47\linewidth]{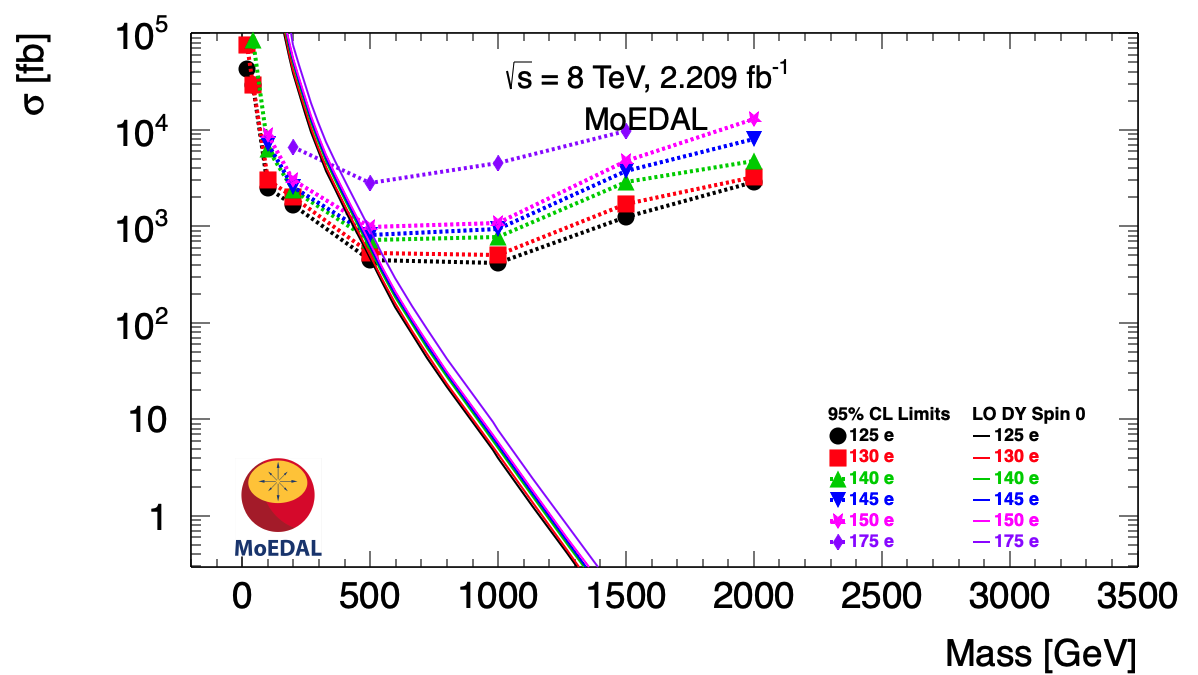}  
    \includegraphics[width=0.47\linewidth]{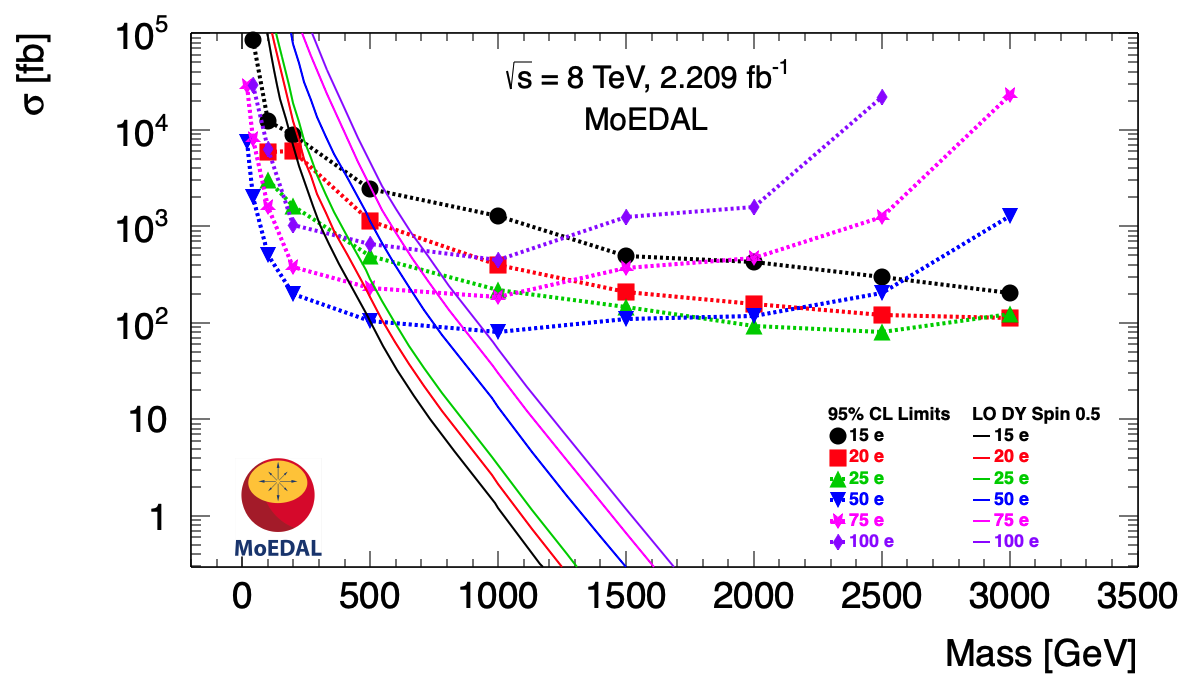}  
    \includegraphics[width=0.47\linewidth]{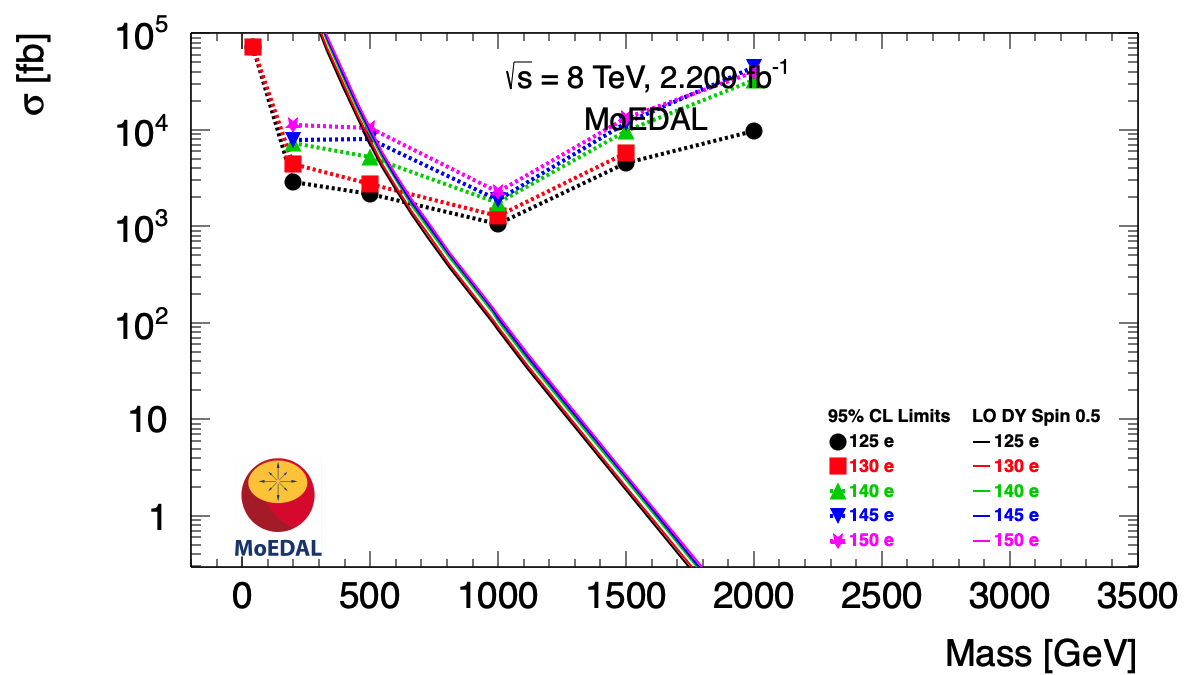}  
    \includegraphics[width=0.47\linewidth]{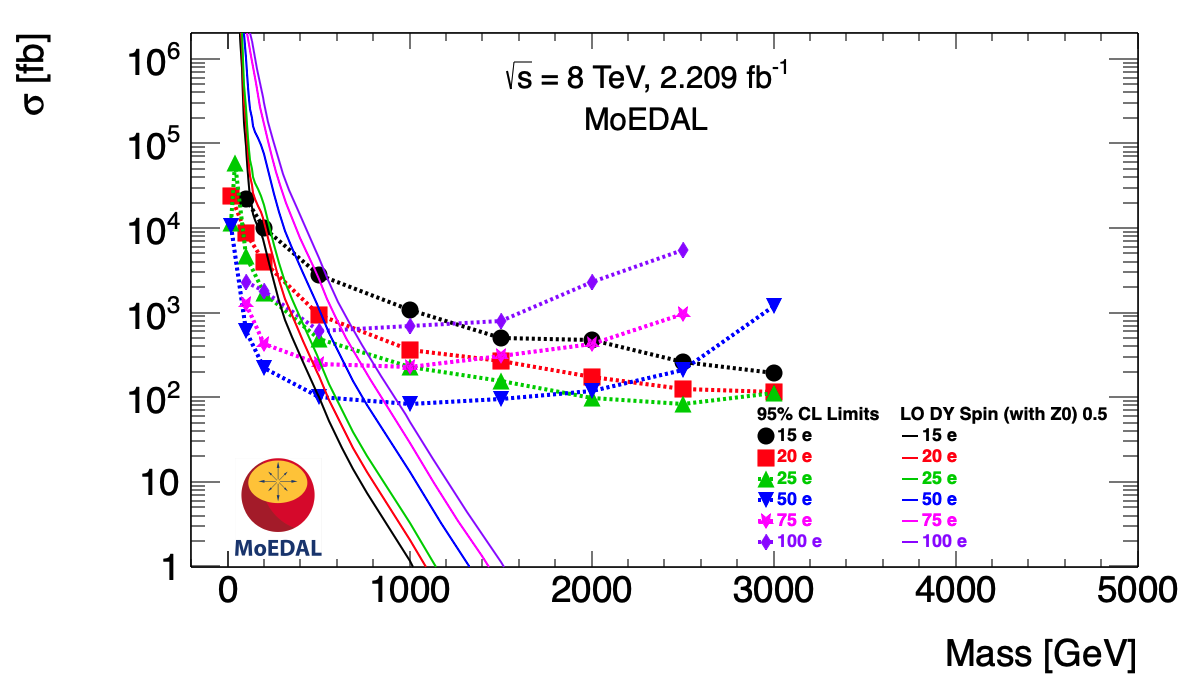}  
    \includegraphics[width=0.47\linewidth]{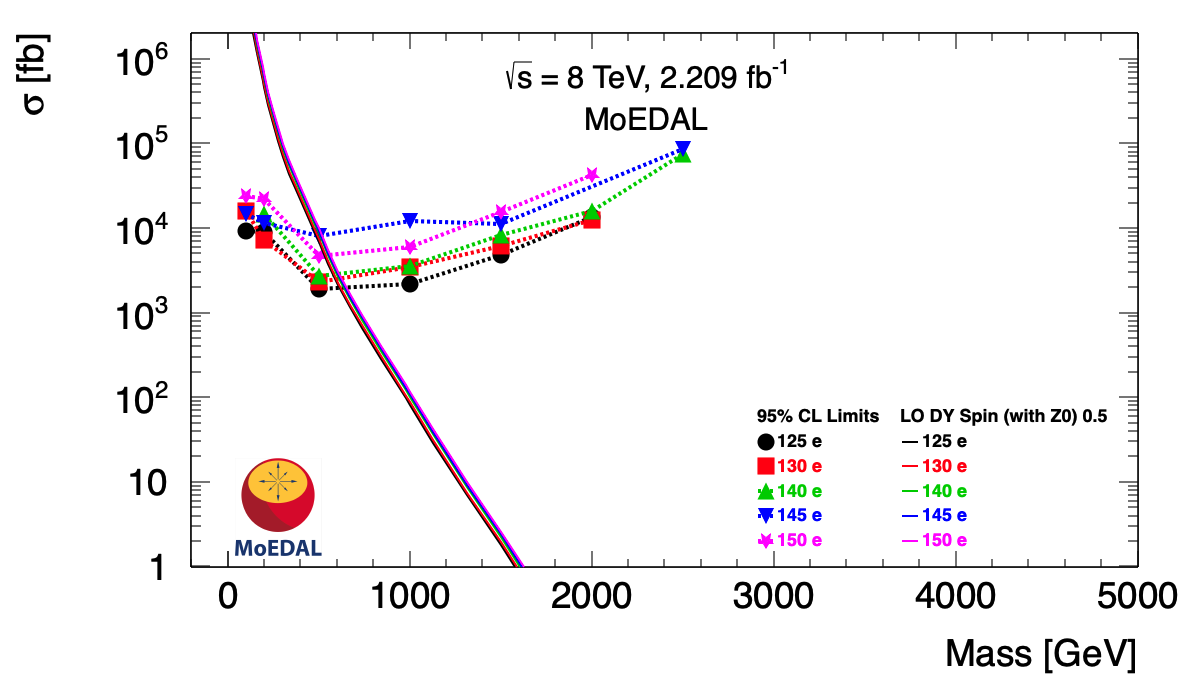}  
      \includegraphics[width=0.47\linewidth]{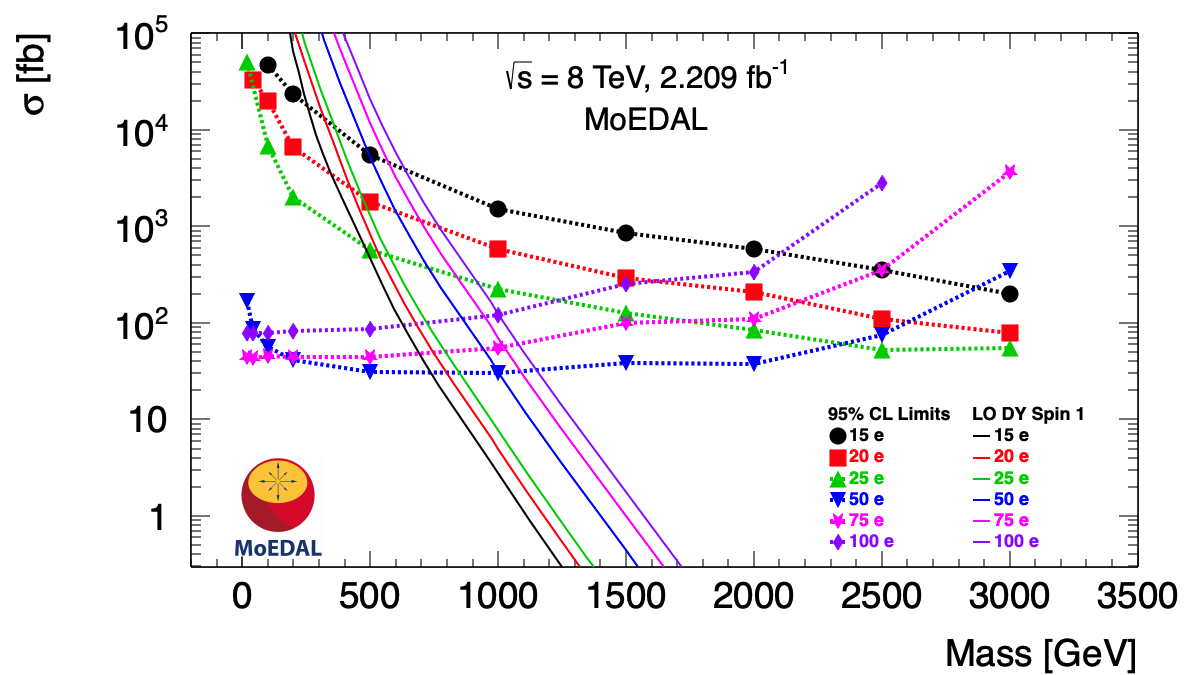}
      \includegraphics[width=0.47\linewidth]{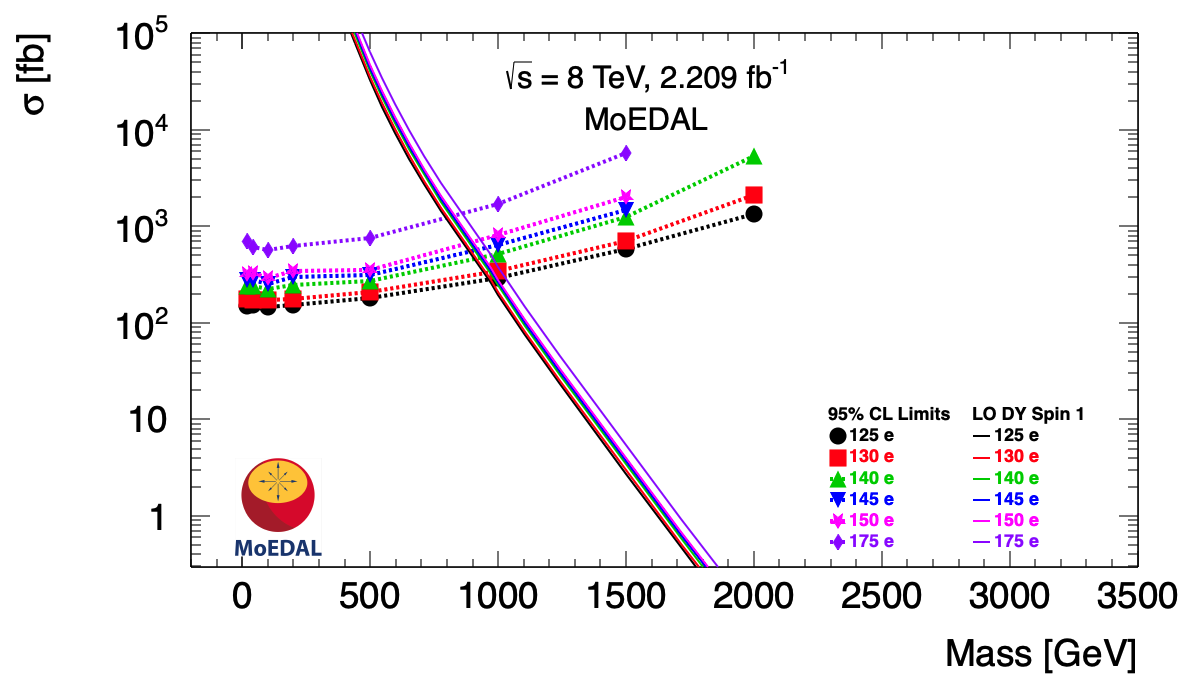} 
    \end{center}
  \caption{\label{fig:limits1} 95\% C.L. upper limits to the cross-section using as a measure a DY model with single virtual photon exchange for spin of $0, 1/2$ and $1$ HECO  production.  Upper limits for the production of spin-1/2 HECOS via DY production including virtual photon and Z$^{0}$ exchange is also included  The solid lines denote the DY cross-sections for each case considered.}
\end{figure*}

\begin{figure*}[!htb]
  \begin{center}
    \includegraphics[width=0.47\linewidth]{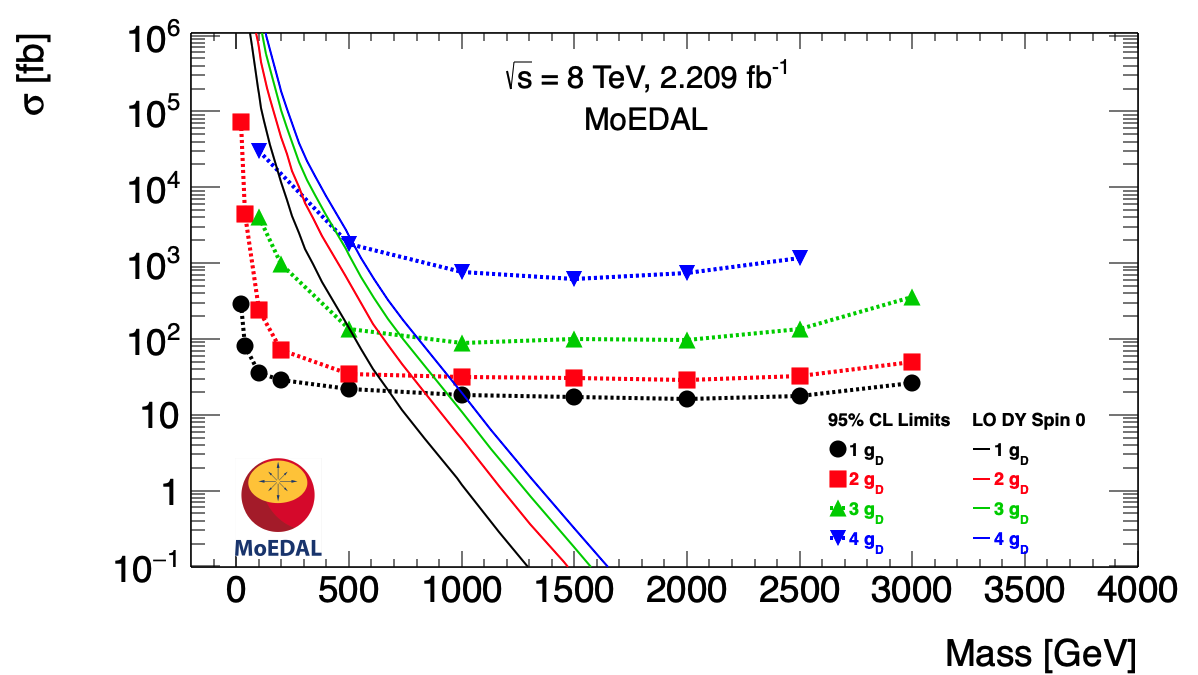}  
    \includegraphics[width=0.47\linewidth]{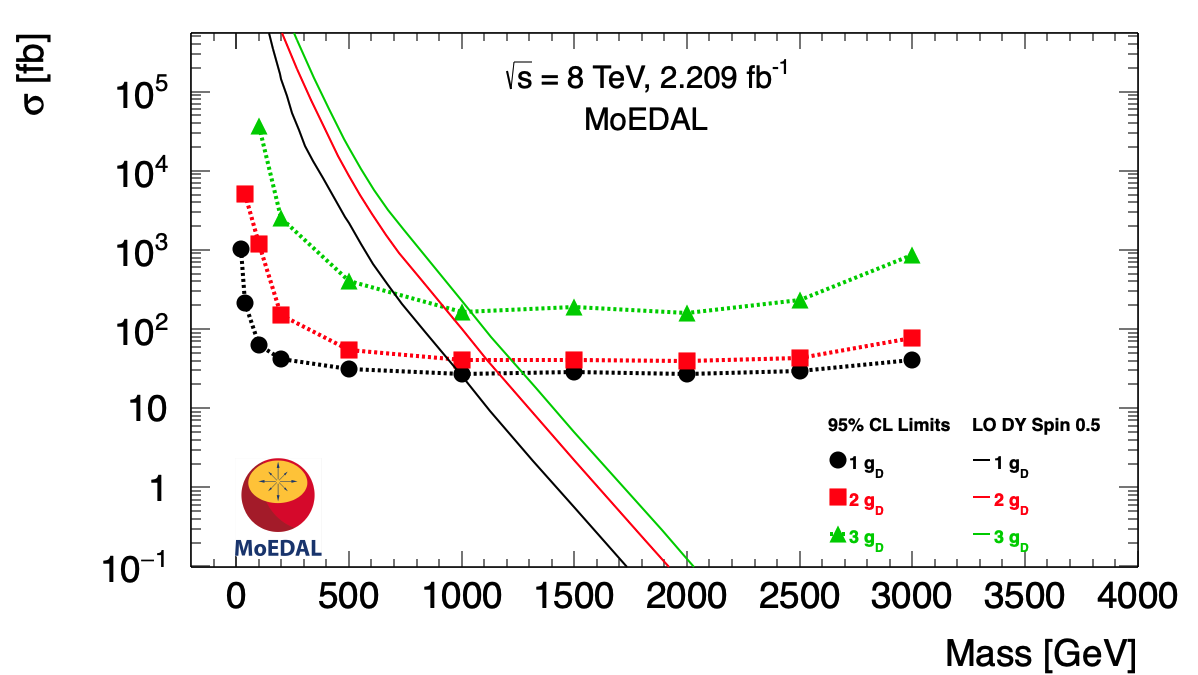}  
    \includegraphics[width=0.47\linewidth]{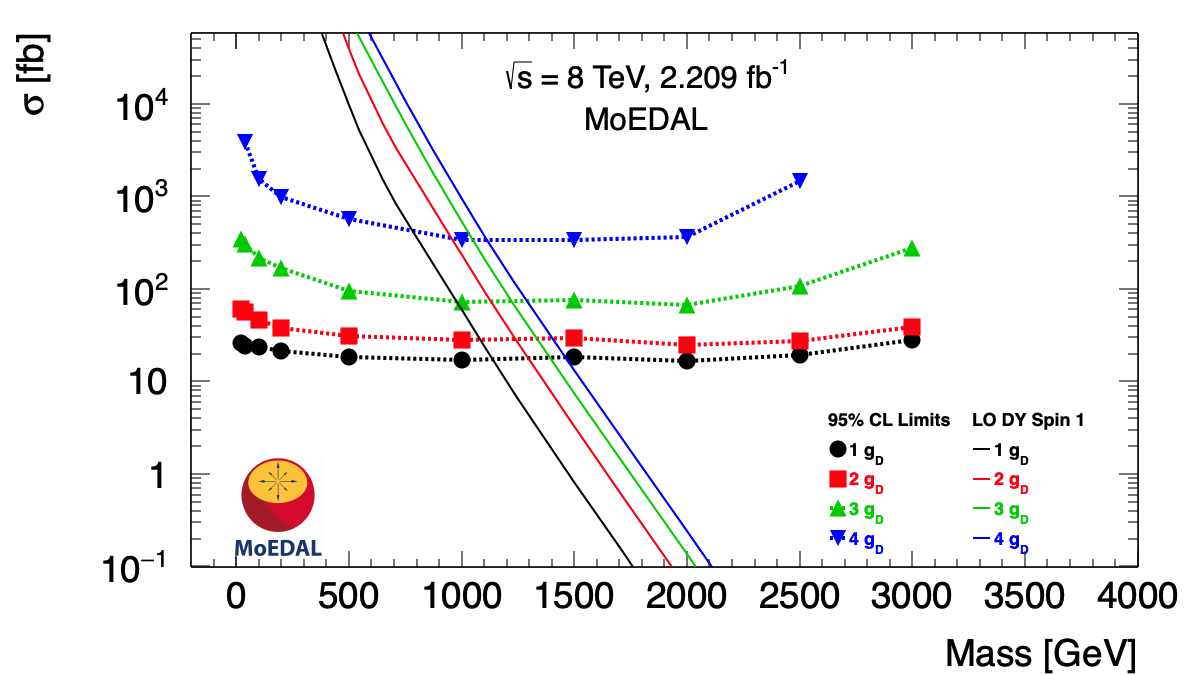} 
    \end{center}
  \caption{\label{fig:limits2}   95\% C.L. upper limits to the cross-section using as a measure a DY model for  spin of $0, 1/2$ and $1$ magnetic monopole production assuming  a $\beta-$independent monopole coupling. The solid lines denote the DY cross-sections for each case considered.}
\end{figure*}

\section{Analysis Results}
The first Makrofol sheet of each of MoEDAL's  125 NTD stacks, exposed during LHC's Run-1 were  etched and scanned, as described above,  for evidence of the passage through the sheet of a highly ionizing object such as a HECO or a magnetic monopole. The total area of plastic analyzed was 7.8~m$^{2}$. No candidate events were observed. In addition, no monopole candidates were observed to be trapped in the MMT detector. This is the first time that the data  from the full MoEDAL prototype detector, deployed during Run-1,  has been presented.

Both the MMTs and the NTDs can be used to detect magnetic monopoles. Consequently, both the NTD and MMT detectors are incorporated into the analysis together. The  Run-1 geometric  acceptance of MoEDAL's NTD and MMT detectors did not overlap. Thus, the procedure  to combine the monopole  signal detected in the  NTD and MMT detectors is  simple addition.
However, only the NTDs can be utilized for the HECO analysis since we have no way to detect electrically charged particles trapped in the MMT detectors. 

The dominant source of systematic error in this analysis arises from  the  imperfect knowledge of the amount of  material 
between the interaction point and  the MoEDAL NTD modules, due to  LHCb's VELO detector.   The VELO vacuum vessel and the various elements of the VELO detector within LHCb's physics acceptance are simulated with great precision in the LHCb geometry.  However, detailed technical drawings of other elements of VELO outside of the physics  acceptance such as cables, in-situ electronics, cooling pipes, various flanges, a vacuum pump and a vacuum manifold,  are not available. 

Nominally,  this intervening material is  between  0.1 and 8.0 radiation lengths ($X_{0}$) in thickness and on average around  1.4 $X_{0}$ 
\cite{LHCb-detector} thick. The  main contribution to the systematic uncertainty in this analysis arises from the estimate of the material in the \geant  geometry description. The uncertainty in the material map is modelled by two geometries which represent an excess  and a deficit of material, using  conservative estimates of  uncertainties on material thicknesses and densities,  compared to the best assessment  of the material budget  that is compatible with direct measurement and existing drawings.

This systematic uncertainty in the material map gives rise to  uncertainties in the DY acceptances. For singly charge monopoles ($|g| = g_{D}$) the resulting relative uncertainty is of the order of 10\%  \cite{Acharya:2016}. This uncertainty increases with  electric and magnetic charge. For a doubly charged monopoles ($|g| = 2g_{D}$)  it is of the order of  10 - 20\% for intermediate masses, around 1 TeV. 

Other sources of systematic error are an uncertainty due to a conservative estimate of  1 cm uncertainty in the trapping detector position. Simulations
show this error lies in  the range 1-17\%  \cite{Acharya:2016}. Another source of systematics is the uncertainty in  $dE/dx$  as a function of $\beta$, resulting in a 1-10\% relative uncertainty in the acceptance \cite{Acharya:2016}.

In the case of monopoles and HECOs a systematic  error on the variable $p$,  due to the NTD etching and calibration process is given in Fig.~\ref{fig:mak-calibs} (bottom). This error on $p$ can give rise to an error on the threshold value for detection of the plastic as well as an error on the variation of efficiency with angle of the NTD. However, these uncertainties are negligible  compared to the error on the material map discussed above.  All of the above sources of systematic error were added in quadrature and included in the final limit calculation. 

We calculated the 95\% C.L. upper limits to the cross-section using as a measure a DY model for HECO and magnetic monopole production assuming  a $\beta-$independent monopole coupling and that the monopole can have a spin of $0, 1/2$ and $1$. The limit curves obtained are shown in Fig.~\ref{fig:limits1} for HECOs. The change in trend of the limit curves as the HECO charge goes above 50$e$ is due to the NTD threshold for detection of highly relativistic HECOs that lies at roughly 50$e$.  For monopoles the cross-section upper limits  versus mass are given in Fig.~\ref{fig:limits2} for spin $0, 1/2$ and $1$.  The values of the corresponding  95\% C.L. mass limits  are  listed in Table~\ref{tab:masslimits-hecos} and Table~\ref{tab:masslimits-monopoles}, for HECOs and magnetic monopoles, respectively. In the case of spin-1/2  HECOs we have  included exclusive DY production limits, both from combined photon/Z$^{0}$ exchange and from simply photon exchange. This allows   our result to be  be compared with the best  published charge limit on HECO production prior to this, from the ATLAS Collaboration \cite{ATLAS-13TeV}, that considered only DY production of HECOs  via photon exchange. 

Note that in the spin-1/2 case and at the mass scale explored in this analysis, the cross-section of DY production of HECOs via photon exchange is slightly larger than the DY HECO production  cross-section via photon/Z$^{0}$ exchange, due  to  destructive interference effects.  The convolution of this lower cross-section with the harder  momentum spectrum and better acceptance of HECOs produced by  DY production via photon/Z$^{0}$ exchange, results in similar  spin-1/2 mass limits  to those obtained from  HECOs produced by DY production via photon exchange only, over most of the  charge range.

 \begin{table*}[!hbt]
\caption{\label{tab:masslimits-hecos} 95\% CL mass limits for the HECO search. }
\vspace{0.5cm}
\begin{tabular}{|c|cccccccccccc|} \hline
       & \multicolumn{12}{c|}{Electric charge (e)}  \\
                                                          &   15    &    20  &  25 &  50    &   75   &  100 &  125&   130 & 140  &145  &  150   &   175  \\ \hline
\colrule
Spin   & \multicolumn{12}{c|}{95\% CL mass limits [GeV/c$^{2}$]  } \\
\colrule
0                                                        &  70   &  120  & 190 &  560  &  580  & 550 & 500  & 490 & 470  & 470  & 460   & 400        \\
1/2 ($\gamma$-exchange)                & 180   &   280 &  440 &  780 &  780   &730  & 660 &  640 & 580  & 520  & 500  &  -                    \\
1/2 ($\gamma$/Z$^{*}$-exchange)   &  170   &  310 &  440 &  780  &  780  &710 &  640 &  620 & 620  & 510  & 580   &  -         \\
1                                                        &  280  &   430 &  590 & 1000 &1020 &1000 & 960 & 950 & 930  &  920  & 900  &  870            \\   \hline
\end{tabular}
\end{table*}


 \begin{table*}[!hbt]\centering
\caption{\label{tab:masslimits-monopoles} 95\% CL mass limits for the magnetic monopole search. }
\vspace{0.5cm}
\begin{tabular}{|c|cccc|} \hline
       & \multicolumn{4}{c|}{Magnetic charge ($\gd$)} \\
       & 1      &   2   &     3  &   4     \\ \hline
\colrule
Spin   & \multicolumn{4}{c|}{95\% CL mass limits [GeV/c$^{2}$]} \\
\colrule
0      & 700    & 790    &   750   &  520          \\  
1/2    & 990   & 1110   &   1040  & -            \\
1     &  1140  & 1240   &   1230 & 1110           \\   \hline
\end{tabular}
\end{table*}

\section{Conclusions}
Both MoEDAL's prototype NTD system and aluminium elements of the MoEDAL MMT detector,  were exposed to 8 TeV LHC  collisions during LHC's  Run-1. At the end of Run-1 both detector systems were examined for the presence of magnetic monopoles and/or HECOs. The NTDs were etched and scanned to reveal evidence for the passage of a magnetic monopole or a HECO  using semi-automatic and manual optical microscopes.  In the case of the MMT  a SQUID-based magnetometer was also utilized to search for  the  presence  of  trapped  magnetic  charge. This is the first time that  search results utilizing the NTD  detectors are presented. 

In previous MoEDAL searches \cite{Acharya:2016} only MoEDAL's  MMT detectors were utilized. Consequently, the HIP search was limited to magnetic monopoles. In this search the use of the NTDs allows the highly ionizing signature of the HIP to be registered. This permits both magnetically charged and electrically charged HIPs (HECOs) to be detected. 

 No magnetic monopole   candidates were  found. Consequently, limits on the DY production of magnetic monopole pair with cross-section in the range of approximately 40~fb to 5~pb were  set  for  magnetic  charges  up  to  4$\gd$ and mass as high  as  1.2 TeV/c$^{2}$. These   limits  are  not competitive  with recent Run-2 collider limits \cite{Acharya:2019,ATLAS-13TeV} despite the use of the NTDs as well as the MMT sub-detectors. This is due to a combination of: the limited acceptance of MoEDAL's  MMT and NTD Run-1 prototype detectors compared to Run-2; the smaller  E$_{CM}$ and DY cross-section at Run-1; and, the smaller luminosity of Run-1 compared to Run-2. 
 
  No evidence was found for DY produced HECO pairs. Thus, limits were placed on the DY production of HECO pairs with cross-sections from around 30 ~fb to 70~pb,  for electric charges in the range 15e to  175e and masses from 110 GeV/c$^{2}$ to 1020 GeV/c$^{2}$. The limits on the DY production of HECOs  are the strongest to date, in terms of charge reach, at any collider experiment.


\section{Acknowledgments}
We thank CERN for the  LHC's successful  Run-1 operation, as well as the support staff from our institutions without whom MoEDAL could not be operated. We acknowledge the invaluable assistance of  particular members of the LHCb Collaboration: G. Wilkinson, R. Lindner, E.  Thomas and G. Corti. In addition we would like to recognize the valuable input from W-Y Song and W. Taylor of York University on HECO production processes.   Computing support was provided by the GridPP Collaboration, in particular by the Queen Mary University of London and Liverpool grid sites. This work was supported by grant PP00P2\_150583 of the Swiss NSF; by the UK Science and Technology Facilities Council, via the grants, ST/L000326/1, ST/L00044X/1, ST/N00101X/1, ST/P000258/1 and ST/T000759/1; by the Generalitat Valenciana via a special grant for MoEDAL and via the projects PROMETEO-II/2017/033 and PROMETEO/2019/087; by MCIU / AEI / FEDER, UE via the grants, FPA2017-85985-P, FPA2017-84543-P and PGC2018-094856-B-I00; by the Physics Department of King's College London; by  NSERC via a project grant; by the V-P Research of the University of Alberta (UofA); by the Provost of the UofA); by UEFISCDI (Romania); by the INFN (Italy); by the Estonian Research Council via a Mobilitas Plus grant MOBTT5;  by a National Science Foundation grant (US) to the University of Alabama MoEDAL group; and, by the National Science Centre, Poland, under research grant 2017/26/E/ST2/00135 and the Grieg grant 2019/34/H/ST2/0070.


\end{document}